\documentclass[notitlepage,tightenlines,aps,pra,reprint,twocolumn,amsmath,amssymb,citeautoscript,nofootinbib,floatfix, longbibliography]{revtex4-1}
\usepackage{float}
\usepackage{graphicx}% Include figure files
\usepackage{dcolumn}% Align table columns on decimal point
\usepackage{bm}% bold math
\usepackage{latexsym,epsfig}
\usepackage{graphicx}
\usepackage{verbatim}
\usepackage{comment}
\usepackage{amsmath} 
\usepackage{amssymb}
\usepackage{stmaryrd}
\usepackage{color}
\usepackage{soul}
\usepackage{epstopdf}
\usepackage{grffile}
\usepackage{mathtools}
\usepackage{physics}
\usepackage[normalem]{ulem}
\usepackage{hyperref}
\hypersetup{
 colorlinks=true, 
 urlcolor=blue,
 citecolor=blue,
 linkcolor=blue
}

\DeclareGraphicsExtensions{.eps}
\newcommand{\beq}{\begin{equation}}
\newcommand{\eeq}{\end{equation}}
\newcommand{\bea}{\begin{eqnarray}}
\newcommand{\eea}{\end{eqnarray}}
\newcommand{\ben}{\begin{eqnarray*}}
\newcommand{\een}{\end{eqnarray*}}
\newcommand{\bfig}{\begin{figure}}
\newcommand{\efig}{\end{figure}}

\begin{document}
\title{Coupling induced emergent topology in a two-leg fermionic ladder}
\author{Rajashri Parida$^{1,2}$, Biswajit Paul$^{1,2}$ , Soumya Ranjan Padhi$^{1,2}$ and Tapan Mishra$^{1,2}$}
\email{mishratapan@gmail.com}
\affiliation{
$^1$School of Physical Sciences, National Institute of Science Education and Research, Jatni,  Odisha 752050, India\\
$^2$Homi Bhabha National Institute, Training School Complex, Anushaktinagar, Mumbai, Maharashtra 400094, India}
\date{\today}

\begin{abstract}
We investigate the ground state properties of spinless fermions on a two leg ladder, by allowing the
nearest-neighbour hopping dimerization in one leg and uniform hopping in the other.
In the non-interacting limit, we find that, at half-filling, the system exhibits robust topological behavior if the inter-leg hopping is allowed. Though depending on the dimerization pattern, the dimerized leg can be either topological or trivial in nature, here we show that by connecting such a leg to a uniform leg through inter-chain coupling, the overall system becomes topological irrespective of the dimerization pattern in the dimerized leg.
As a result, a topological phase transition occurs as a function of the inter-leg hopping. When the inter-leg interaction is turned on, the topological phase survives, and we obtain an interaction induced topological phase transition. 
Finally, we reveal that when uniform interactions are included on all the bonds of the ladder, the topological phase transitions to a symmetry-broken charge-density wave (CDW) phase.
\end{abstract}

\maketitle
\section{Introduction}
Over the past decades, the notion of topology has emerged as a powerful framework for understanding novel quantum phases of matter that fall beyond the conventional Landau symmetry-breaking paradigm~\cite{wilson_phase_transition}. 
Unlike conventional phases classified by local order parameters and spontaneous symmetry breaking, topological phases are characterized by global, quantized invariants associated with the wavefunctions of the system. 
Band topology, in particular, has reshaped the classification of insulating phases in non-interacting systems, giving rise to robust edge or surface states protected by underlying symmetries~\cite{Hassanreview}. 
One such example is the symmetry-protected topological (SPT) phase, which possesses robust boundary states as long as certain symmetries are preserved.
The celebrated model that exhibits band topology and an SPT phase is the Su-Schrieffer-Heeger (SSH) model of non-interacting fermions~\cite{ssh_model}, which hosts topologically protected edge states arising from the dimerized hopping in a one-dimensional chain. 
The SSH model, due to its simplicity, has been extensively studied~\cite{Asboth2016_ssh,ssh_3,ssh_su_chen,gloria_platero_ssh,gen_ssh}, and its topological features have been realized in diverse experimental platforms, including cold atoms, photonic and acoustic lattices, superconducting and electrical circuits, ion traps, and mechanical systems~\cite{Atala2013,Takahashi2016pumping,Lohse2016,Mukherjee2017,Lu2014,ssh_expt_1,ssh_expt_2, ssh4_expt,Kitagawa2012,Leder2016,browyes,semicoductor_nanolattice,seba_soliton,Liu-superconducting,rydberg_atom_review, EC1, EC2}.

These ideas have now expanded beyond non-interacting limits, opening new avenues for exploring how strong correlations, dimensionality, and symmetry intertwine to stabilize or destabilize topological order in the free fermionic counterpart~\cite{rachel_review}.
In this context, SPT serves as the simplest playground to study the interplay of these key aspects of quantum condensed matter physics, as these phases are short-range
entangled phases~\cite{chen_gu_wen,senthil_review}.
Due to this, extensive research has been conducted using interacting SSH models to understand how local and non-local interactions affect topological phases and phase transitions in many-body fermionic, bosonic, and spin systems~\cite{fleishauer_prl, DiLiberto2016,DiLiberto2017,sjia, wu_vincet_liu, Taddia2017,fraxanet,juliafare_pumping, mondal_sshhubbard,pedro_ssh,XXZ_chain_Aligia,pan_interaction_induced,mohamadi_ssh}. 
An interesting revelation in this context is the interaction induced topology where interactions alone give rise to nontrivial topology in systems that are otherwise topologically trivial in the non-interacting limit~\cite{Mondal_topology,hetenyi_tvvp,juliafarre_efh,topo_end_states, parida_topology,julia_fare_3}.

On the other hand, interaction effects and dimensionality of the lattices on the topological phases have been investigated by going beyond strictly one-dimensional chains. 
In particular, ladder geometries or quasi-one dimensional lattices that are intermediate to one and two-dimensional lattices have attracted a great deal of attention, revealing interesting topological properties ~\cite{Nersesyan,ssh_ladder_glide_symmetry,ssh_ladder_tnp,smita_ssh_ladder,padhan_ladder, suman_adhip, batrouni,wu_vincet_liu,rajashri_ladder,unconventional_edge_state,four_spin_furukawa,frustrated_xxz_furukawa,spt_spin_ladder,sara_ladder,thereza_haldane,floquet_ssh_ladder,milad_ssh_ladder}. 
The key component in the ladder systems that plays the major role is the inter-layer coupling that decides the fate of the topological properties of individual chains and may impart novel physical phenomena due to proximity effects~\cite{semiconductor_metal_proximity,topo_proximity_effect,btpe,haldane_graphene_bilayer,hofstetter_prximity,uneven_ladder,uneven_ladder_2,bowtie_ladder}.
For instance, Refs.~\cite{nersesyan_ssh_ladder,smita_ssh_ladder,ssh_ladder_tnp} investigate the emergence of topological phase transitions when two SSH chains are coupled through inter-chain hopping, particularly in situations where one leg is topological while the other is trivial. Interaction effects in such ladder systems have also been explored in Refs.~\cite{nersesyan_ssh_ladder,padhan_ladder}.
Furthermore, the topological inheritance in a two-leg ladder through inter-leg interaction has been predicted in Ref.~\cite{mondal_sshhubbard}. It has been shown that an uniform chain can inherit the topological character of an SSH chain when inter-chain interaction is tuned and in the absence of any inter-chain hopping.
Apart from the simple two-leg ladder, the topological properties of different ladders with non-standard interleg couplings have been explored recently.
In this context, Ref.~\cite{bowtie_ladder} analyzes coupled tight-binding chains and demonstrates that the emergence of topological features is sensitive to the choice of unit cell, with nontrivial phases arising only for the choice of unit cell that results in dangling edge sites under open boundary conditions. 
On the other hand, studies of uneven ladders composed of chains with different periodicities~\cite{uneven_ladder,uneven_ladder_2} show that the absence of chiral symmetry leads to edge states that are not pinned to zero energy, and hence the topological insulator here does not belong to the BDI class. 
Importantly, as emphasized in Ref.~\cite{bowtie_ladder}, when two identical tight-binding chains are coupled via simple vertical inter-chain hopping without any mismatch in periodicity, no topological features arise.
These studies suggest that the two-leg ladder platforms provide a fertile ground to explore novel topological scenarios in low-dimensional lattices.

Motivated by this, in this work, we study the topological properties of an asymmetric two-leg ladder model composed of spinless fermions, where one leg follows the SSH-type dimerized hopping, and the other is a uniform tight-binding chain. 
For the non-interacting case, the dimerization pattern decides the topological or trivial nature of the SSH-type leg. However, once coupled to a uniform tight-binding chain through interchain tunneling, the ladder as a whole becomes topological regardless of the dimerization pattern in the dimerized leg.
Interestingly, in the many-body limit,  including the rung interaction, it also drives a topological phase transition in the system as a function of rung interaction.
However, when uniform interactions are introduced along both the legs and rungs, this topological phase transition ceases to happen, and the system instead undergoes a transition into a CDW phase.

The paper is organized as follows. 
In sec.~\ref{modelmethod}, we introduce the model under consideration and the computational method used to obtain all the results. 
In sec.~\ref{noninteracting}, we present the non-interacting case, focusing on both regimes of dimerization in the SSH-type leg. 
Sec.~\ref{many_body} discusses the scenario when interaction between the particles is considered.
Finally, in sec.~\ref{conclusion}, we provide a summary of the key results.

\section{Model}\label{modelmethod}
\begin{figure}[t]
\begin{center}
\includegraphics[width=1\columnwidth]{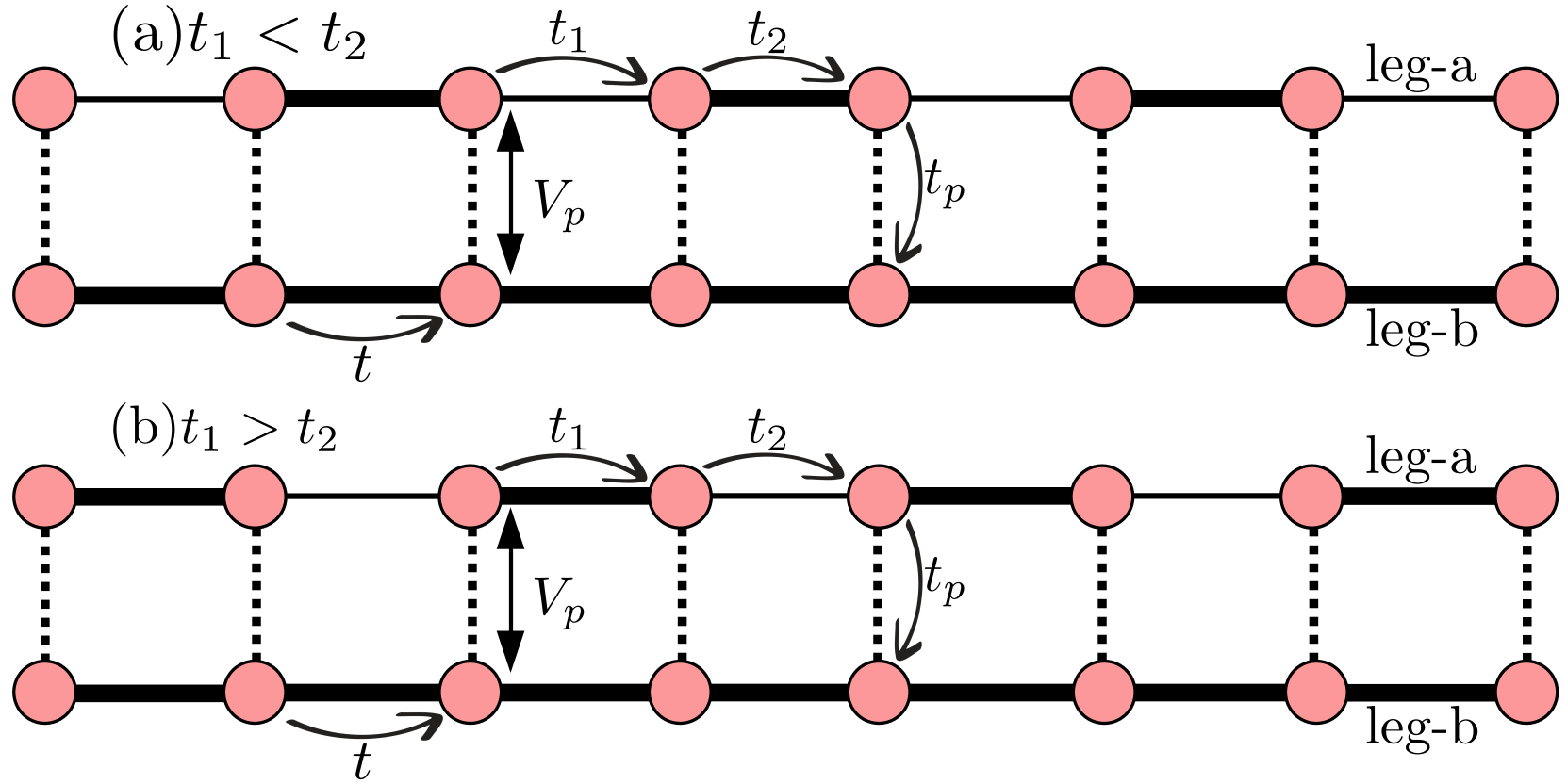}
\end{center}
\caption{Schematic of a two-leg ladder with dimerized hopping along the upper leg. The circles denote lattice sites. The alternating hopping amplitudes are $t_1$ and $t_2$, where the stronger hopping is represented by thick solid lines and the weaker hopping by thin solid lines. The lower leg has uniform hopping, shown by thick solid lines. Panel (a) corresponds to $t_1<t_2$, while panel (b) corresponds to $t_1>t_2$.}
\label{fig:schematic}
\end{figure}

We consider a system of spinless fermions on a two-leg ladder depicted in Fig.~\ref{fig:schematic}, where uniform and SSH type dimerized hopping patterns in the lower and the upper legs have been assumed, respectively. 
The model that describes such a system is given by the Hamiltonian,
\begin{align}
    \hat{H} &= -\bigg{(}\sum_{i\in\text{odd}}t_1\hat{a}_i^\dagger\hat{a}_{i+1}+ \sum_{i\in\text{even}}t_2\hat{a}_i^\dagger\hat{a}_{i+1} +t\sum_i\hat{b}_i^\dagger \hat{b}_{i+1} \nonumber \\
    &\;+t_p\sum_i\hat{a}_i^\dagger \hat{b}_i\bigg{)}+\text{H.c.}
    \label{eq:single_particle_ham}
\end{align}
$\hat{H}$ represents the Hamiltonian of non-interacting spinless fermions.
$\hat{a}_i(\hat{a}_i^\dagger)$ and $\hat{b}_i(\hat{b}_i^\dagger)$ are the particle annihilation (creation) operators on leg-a and leg-b, respectively, where $i$ represents the rung index. $t_1$ and $t_2$ are the alternating hopping amplitudes in the upper leg, $t$ is the uniform hopping strength in the lower leg, and $t_p$ represents the strength of the rung (inter-leg) hopping. Throughout all the numerical calculations, we set $t=1$, which sets the energy scale of the system.

Since the single-particle Hamiltonian $\hat{H}$ has a unit cell consisting of four lattice sites, Fourier transforming the Hamiltonian in Eq.~\eqref{eq:single_particle_ham} yields
\begin{equation}
    \hat{H}_k = \begin{bmatrix}
    0 & \mu_1 & 0 & -t_p\\
    \mu_1^* & 0 & -t_p & 0\\
    0 & -t_p & 0 & \mu_2\\
    -t_p & 0 & \mu_2^* & 0\\ 
    \end{bmatrix}
    \label{eq:band_ham}
\end{equation}
where we have set the lattice constant to unity, $\mu_1=-t_1-t_2e^{-ik}$ and $\mu_2=-t(1+e^{ik})$, with $k$ representing the quasi-momentum.

In terms of Pauli matrices, the above Hamiltonian can be written as 
\begin{eqnarray}
    \hat{H}_k &=& -\{[(t_1+t)+(t_2+t)\text{cos}(k)]\sigma_x\nonumber\\&+&
    (t_2-t)\text{sin}(k)\sigma_y\}\otimes \frac{\mathbf{I}}{2}\nonumber\\&-&\{[(t_1-t)+(t_2-t)\text{cos}(k)]\sigma_x\nonumber\\&+&
    (t_2+t)\text{sin}(k)\sigma_y\}\otimes \frac{\nu_z}{2}-t_p \sigma_x\otimes\nu_x
    \label{eq:pauli_matrix_form}
\end{eqnarray}
where $\nu_m$'s and $\sigma_m$'s with $m=(x,~y,~z)$, are the Pauli matrices corresponding to the leg of the ladder and sublattice in each leg. 
$\mathbf{I}$ represents the $2\times2$ identity matrix.

\begin{figure*}
\centering
\includegraphics[width=2.\columnwidth]{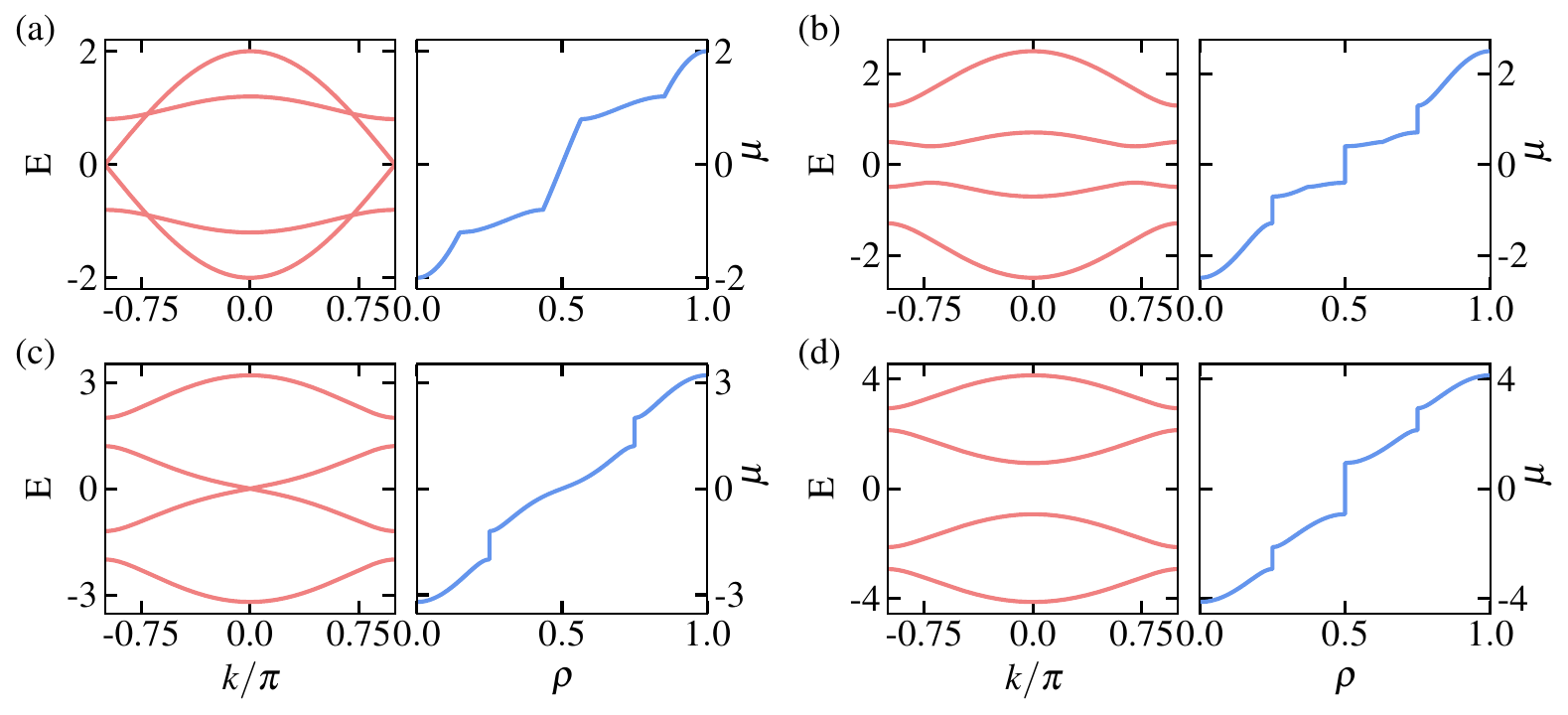}
\caption{The left panels in (a-d) represent the single particle dispersion bands for $t_p=0.0,~0.8,~\sqrt{2.4},~\text{and}~2.5$ respectively. The right panels in (a-d) represent the corresponding particle filling ($\rho$) plotted against the chemical potential ($\mu$) for the same values of $t_p$ with a system of $L=500$ rungs. For all the plots, the values of $t_1$ and $t_2$ are fixed at $0.2$ and $1.0$ respectively.}
\label{fig:band_rho_mu}
\end{figure*}

In the following, we discuss our main results.
\section{Results}\label{results}
In this section, we first analyze the noninteracting case, where we investigate the effect of rung hopping, and then we investigate the role of interparticle interaction.
%%%%%%%%%%%%%%%%%%%%%%%%%%%%%%%%%%%%%%%%%%%%%%%%%%%%%%%%%%%%%%
\subsection{Non-interacting case}\label{noninteracting}
%%%%%%%%%%%%%%%%%%%%%%%%%%%%%%%%%%%%%%%%%%%%%%%%%%%%%

To understand the topological nature of the non-interacting particles, we begin by examining the gap-opening as well as the gap-closing scenarios.
Since in fermionic systems, the energy gaps can be fully understood from the underlying band structure, we start by diagonalizing the Hamiltonian given in Eq.~\eqref{eq:pauli_matrix_form}, and the resulting dispersion relation is
\begin{align}
    &E^2=t_p^2 + \frac{1}{2}(|\mu_1|^2+|\mu_2|^2)\nonumber\\&\pm\frac{1}{2}\sqrt{(2t_p^2+|\mu_1|^2+|\mu_2|^2)^2+4(\mu_1\mu_2-t_p^2)(t_p^2-\mu_1^*\mu_2^*)},\nonumber\\
    \label{eq:dispersion}
\end{align}
where, $\mu_1$ and $\mu_2$ are as defined earlier. 

The above dispersion relation suggests that for any finite dimerization in hopping in leg-a, the gap closes at $k=0$ when $t_p^2=2t(t_1+t_2)$, and at $k=\pi$, when $t_p^2=0$.
At all other points, the gap remains open at half-filling. 
We display the band structure for four exemplary values of $t_p$: $t_p = 0$, $0.8$, $\sqrt{2.4}$, and $2.5$ in the left panles of Fig.~\ref{fig:band_rho_mu}(a–d) respectively by fixing $t_1=0.2$ and $t_2=1.0$.
To compare the band picture with the filling of particles in real space, the filling $\rho = N/2L$ as a function of chemical potential $\mu$~\cite{rmi_laflorencie} calculated under PBC, is plotted in right panels of Fig.~\ref{fig:band_rho_mu}(a–d), where $N$ is the total number of fermions and $L$ is the number of rungs.
The appearance of cusps in the $\rho$–$\mu$ curves reflects changes in the number of Fermi points i.e., the number of quasi-momenta where the Fermi energy intersects the bands.
The width of each plateau in the $\rho$–$\mu$ curve indicates the value of the energy gap at the corresponding filling, whereas the shoulder regions around the plateaus indicate the gapless Luttinger liquid.
Comparing the two panels in Fig.~\ref{fig:band_rho_mu}(a), we observe that for $t_p = 0$, the system remains gapless at all fillings. 
In contrast, for $t_p = 0.8$, finite gaps emerge at $\rho=1/4,~1/2,~\text{and}~3/4$, as shown in Fig.~\ref{fig:band_rho_mu}(b). 
Increasing the value of $t_p$ to $\sqrt{2.4}$ closes the gap at $\rho=1/2$, while the gaps at $\rho=1/4,~\text{and}~3/4$ persist [see Fig.~\ref{fig:band_rho_mu}(c)]. 
At $t_p = 2.5$, gaps reappear again at all three commensurate fillings, such as for $\rho=1/4,~1/2,~\text{and}~3/4$, as shown in Fig.~\ref{fig:band_rho_mu}(d).

With this information in hand about the gap of the system, we investigate topological phase transitions in the subsequent subsection.

\subsubsection{Topological Phase Transition}%
\begin{figure}[t!]
    \centering
    \includegraphics[width=1\columnwidth]{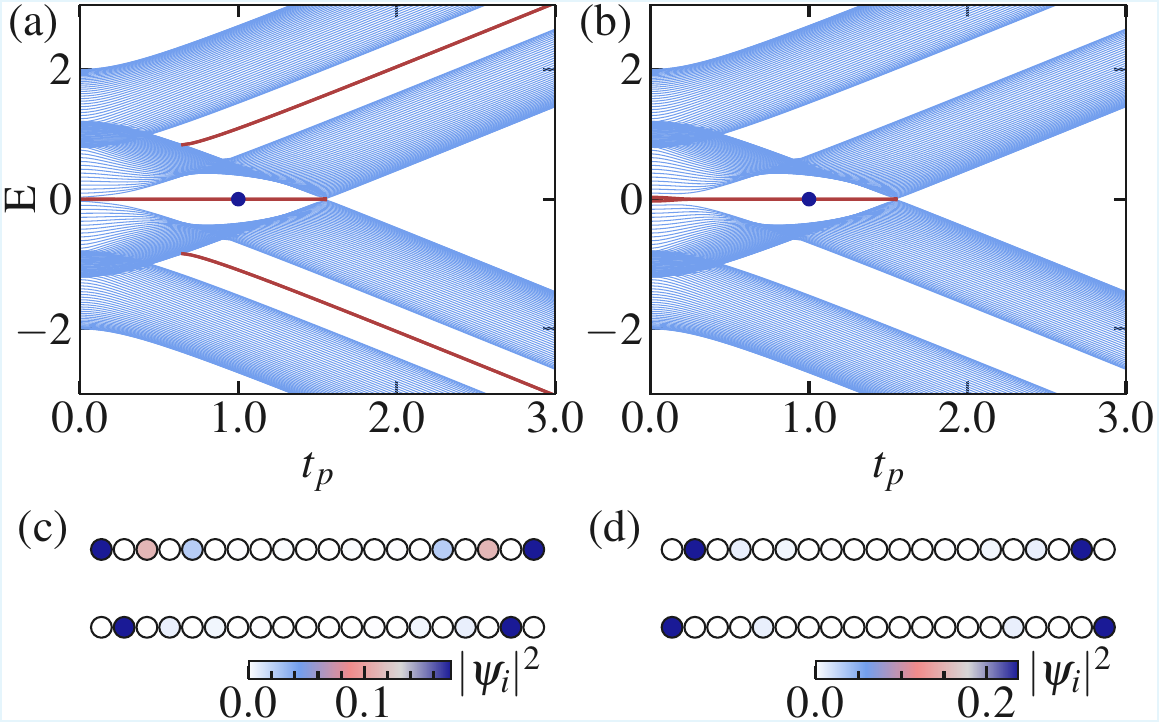}
    \caption{(a) and (b) energy spectrum as a function of rung hopping $t_p$ when $t_1=0.2$, $t_2 = 1.0$ and $t_1=1.0$, $t_2 = 0.2$ respectively. (c) and (d) represent the probability density $|{\psi_i}|^2$, where the circles represent the sites of the ladder and the facecolor of the circle represents the probability density corresponding to the blue circle in (a) and (b), respectively.}
    \label{fig:E_spec}
\end{figure}

Since one of the legs of the ladder is considered to be of SSH type, we can have two possible dimerization patterns: (i) topological pattern ($t_1<t_2$ as in Fig.~\ref{fig:schematic}(a)) and (ii) the trivial pattern ($t_1>t_2$ as in Fig.~\ref{fig:schematic}(b)). For the topological (trivial) pattern, we choose $t_1=0.2,~t_2=1.0$ ($t_2=1.0,~t_2=0.2$). 
Note that the hopping in the other leg is uniform throughout and is set to $t=1.0$. This choice of hopping suggests that in the decoupled leg limit, the leg-a can be either trivial or a topological insulator whereas the leg-b is always a gapless Luttinger liquid. Our goal is to investigate the influence of the dimerized leg on the system when the two legs are coupled to each other. 
We first explore the real space energy spectrum of the system under OBC. In Fig.~\ref{fig:E_spec}(a) and (b) we depict the energy spectrum as a function of $t_p$ for $t_1<t_2$ and $t_1>t_2$ configurations, respectively.  
For both the configurations, the bulk gaps open up as soon as $t_p$ becomes finite. Interestingly, for both the topological and trivial dimerization configuration of leg-a, the bulk gap hosts two mid-gap zero-energy states which disappear at a critical $t_p=\sqrt{2.4}$ with the closing up of the bulk gap. However, the gap reopens with further increase in $t_p$ without the zero energy states. These zero energy states are typical signatures of topological phases under OBC.
In order to examine that these zero-energy states are edge states, we plot the onsite probability density ($|{\psi_i}|^2$) corresponding to one of these two states.
As shown in Fig.~\ref{fig:E_spec}(c) and (d) the states corresponding to the dimerization patterns $t_1=0.2,~t_2=1.0$ and $t_1=1.0,~t_2=0.2$ for an exemplary value $t_p=1.0$ (dots in Fig.~\ref{fig:E_spec}(a) and (b)), are localized at the edges of the system. 
The key distinction between both the cases is that the edge states exist on the upper leg when $t_1<t_2$ and on the lower leg when $t_1>t_2$.
Interestingly, we observe the lower-leg, which doesn't have any hopping dimarization, has supported a zero energy edge state.
Additionally, the spectra also exhibit isolated energetic mid-gap states at $\rho=1/4$ and $\rho=3/4$ fillings for $t_1<t_2$ (the red solid lines), which are absent for $t_1>t_2$.
For $\rho=1/2$, it is evident that there is a pair of localized edge modes existing in the upper leg for the $t_1<t_2$ case, whereas at $\rho=1/4$, the edges of both the legs of the ladder hybridize (not shown), resulting in energetic edge states. 
In contrast, there is also another pair of edge modes at zero energy for $t_1>t_2$ case as shown in Fig.~\ref{fig:E_spec}(d), and these are localized in the lower leg of the ladder. 
From now on, we focus on the zero-energy edge states appearing at half-filling to characterize the topological property of our system, as these are the states protected by symmetry.

\begin{figure}[t]
\begin{center}
\includegraphics[width=1\columnwidth]{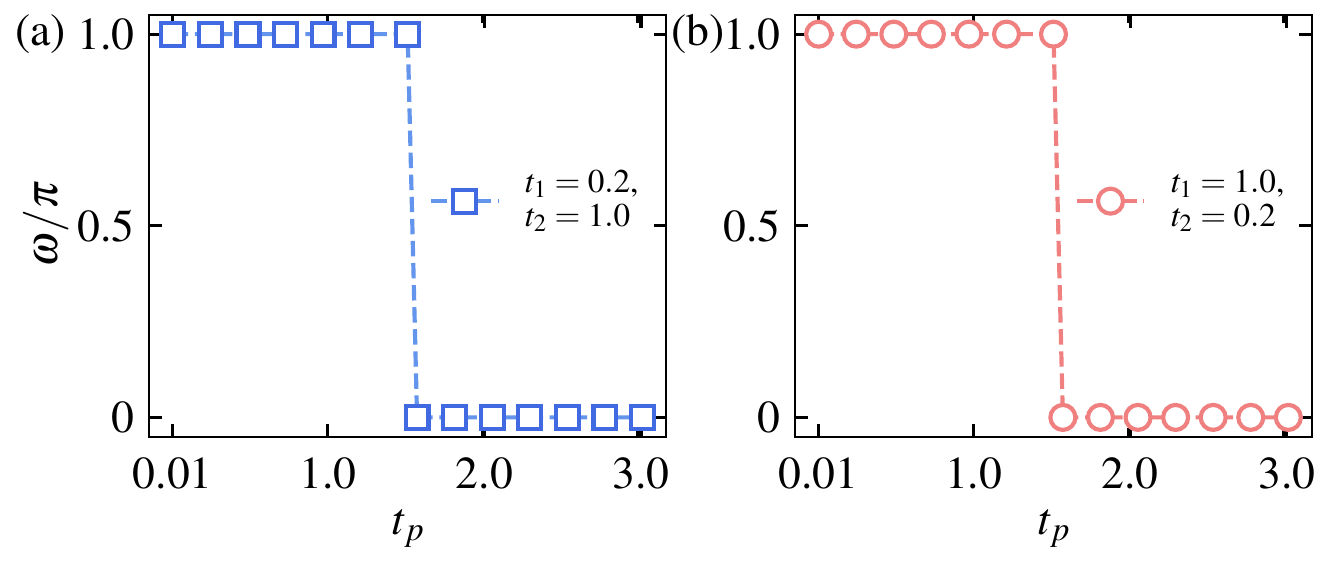}
\end{center}
\caption{(a) and (b) illustrate winding number as a function of $t_p$ when $t_1=0.2$, $t_2=1.0$ and $t_1=1.0$, $t_2=0.2$, respectively.}
\label{fig:berry_phase}
\end{figure}

To concretely establish this topological feature, we compute the topological invariant called the Berry phase ($\omega$)~\cite{berry_phase}, which is defined as 
\begin{equation}
  \omega = -\mathfrak{I}\ \ln \prod_{j=0}^{N-1} \text{det}(\langle u_{m,k_j} |u_{n,k_{j+1}} \rangle),
  \label{eq:berry_phase}
\end{equation}
where, $\mathfrak{I}$ denotes the imaginary part, $k_j=2\pi j/Na$ is the $j^{\text{th}}$ $k$ point in the Brillouin zone and $|u_{m,k_j}\rangle$ denotes the occupied Bloch bands with $m$ and $n$ representing the band indices.
The determinant is taken as we are dealing with a multiband system.
It is well known that a trivial bulk is characterized by $\omega=0$, while a topological bulk possesses a value $\omega=\pi$. 
We calculate the Berry phase $\omega$ for both $t_1<t_2$ as well as $t_1>t_2$ cases by taking the filled bands up to half filling and plot $\omega/\pi$, called the winding number, as a function of $t_p$ in Fig.~\ref{fig:berry_phase}(a) and (b), respectively. 
At $t_p=0$, the two legs of the ladder are completely decoupled, with one forming a tight-binding chain for which the gap closes at $k=\pi$, making the Berry phase undefined. 
For any finite $t_p$, however, the system acquires a well-defined Berry phase, which takes the value $\pi$ for both $t_1<t_2$ and $t_1>t_2$ for arbitrarily small $t_p$, confirming the topological nature of the phases. 
As $t_p$ increases, the bulk gap closes at $t_p=\sqrt{2.4}$, beyond which the Berry phase drops to $0$, indicating a transition to a trivial phase. 
This change in $\omega/\pi$ from 
$1$ to $0$ for both $t_1<t_2$ and $t_1>t_2$ provides clear evidence of a topological phase transition at $t_p=\sqrt{2.4}$. 
Furthermore, this transition is also accompanied by the disappearance of the zero-energy edge modes present in the topological phase, and this is called the bulk-boundary correspondence in our system.

\begin{figure}[t]
\begin{center}
\includegraphics[width=1\columnwidth]{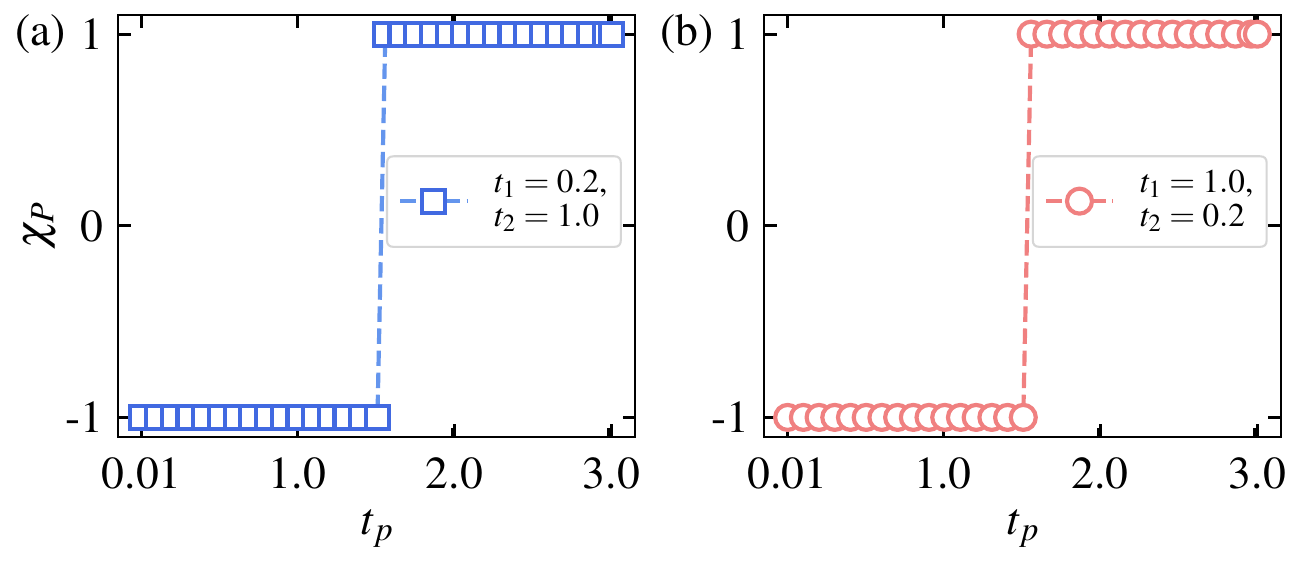}
\end{center}
\caption{Variation of $\chi_{P}$ as a function of $t_p$ for (a) $t_1=0.2,~t_2=1.0$ and (b) $t_1=1.0,~t_2=0.2$ respectively.}
\label{fig:chiP}
\end{figure}

In addition to the above analysis, we also explain the origin of topology in our system from the perspective of symmetry, for which we rely on the inversion symmetry present in our system.
The Bloch Hamiltonian satisfies
\begin{equation}
\hat{\mathcal P}\,\hat H(k)\,\hat{\mathcal P}^{-1} = \hat H(-k),
\end{equation}
where the inversion operator is given by
\begin{equation}
\hat{\mathcal P} =
\begin{bmatrix}
0 & 1 & 0 & 0\\
1 & 0 & 0 & 0\\
0 & 0 & 0 & 1\\
0 & 0 & 1 & 0
\end{bmatrix}.
\label{eq:inv_op}
\end{equation}
At the inversion-symmetric momenta $k_{\mathrm{inv}} = 0,\pi$, the Hamiltonian commutes with $\hat{\mathcal P}$, and the Bloch eigenstates can therefore be chosen as simultaneous eigenstates of inversion. Each occupied band is characterized by an inversion eigenvalue $\xi_i(k_{\mathrm{inv}})=\pm 1$.
The topology of the occupied bands is characterized by a quantity $\chi_P$ defined as
\begin{equation}
\chi_P = \prod_{k_{\mathrm{inv}}=0,\pi}\;\prod_{i \in \mathrm{occ}} \xi_i(k_{\mathrm{inv}}).
\label{eq:chiP}
\end{equation}
For our case, $i$ will go over the lowest two bands as we are discussing the situation at half-filling.
In practice, the inversion eigenvalues are obtained by projecting the inversion operator onto the occupied subspace at each inversion-symmetric momentum point. 
If $\hat{P}_k$ denotes the projector onto the occupied bands at momentum $k$, the inversion eigenvalues are given by
\begin{equation}
\hat{P}_k\,\hat{\mathcal P}\,\hat{P}_k,
\qquad k=0,\pi.
\end{equation}
A change in the parity structure of the occupied bands between $k=0$ and $k=\pi$ signals a nontrivial twist, whereas identical inversion eigenvalues at both momenta correspond to a topologically trivial phase.
For example, for $(t_1,t_2,t_p)=(0.2,1.0,0.2)$ one finds
$\{\xi_1(0),\xi_2(0)\}=(-1,-1)$ and $\{\xi_1(\pi),\xi_2(\pi)\}=(1,-1)$, giving $\chi_P= \xi_1(0)\xi_2(0)\xi_1(\pi)\xi_2(\pi) =-1$.
For $(t_1,t_2,t_p)=(0.2,1.0,2.0)$,
$\{\xi_1(0),\xi_2(0)\}=(-1,1)$ and $\{\xi_1(\pi),\xi_2(\pi)\}=(1,-1)$, yielding $\chi_P=+1$.
Similarly, for $(t_1,t_2,t_p)=(1.0,0.2,0.2)$ one obtains $\chi_P=-1$.
We calculate the value of $\chi_p$ as a function of $t_p$ for both $t_1<t_2$ and $t_1>t_2$ cases and show in Fig.~\ref{fig:chiP}(a) and (b) of this response, respectively.
The change of sign of $\chi_p$ from $-1$ to $1$ at critical $t_p$ supports the topological phase transition happening in the system.

Now, to explain the origin of topology for both possible dimerization configurations, we derive effective Hamiltonians for the individual legs of the ladder, in the weak inter-leg coupling limit, $t_p \ll t_1,t_2$. 
As shown in Fig.~\ref{fig:E_spec}(c) and (d), the location of the zero-energy edge modes depends on the dimerization pattern. 
While for $t_1<t_2$, the edge modes are localized on the leg-a, for the $t_1>t_2$ case, these modes are located in the leg-b. 
The behaviour for the $t_1<t_2$ case is expected because the leg-a, when considered independently, corresponds to the topological phase of the SSH model~\cite{ssh_model}. 
Consequently, the topological character of the full ladder is inherited from the leg-a. 
Since the complete system preserves chiral symmetry, the remaining couplings only renormalize the properties of an already existing topological phase.
However, the situation is markedly different for $t_1>t_2$. 
In this case, the zero-energy edge modes are found on the leg-b, even though neither of the legs is individually topological. 
While the leg-a corresponds to a trivial SSH configuration, the leg-b is simply a gapless tight-binding chain. 
Therefore, the emergence of edge states and the associated topological bulk behaviour cannot be understood from the properties of the isolated legs alone. 
Instead, the inter-leg coupling induces an effective renormalization of the hopping amplitudes, giving rise to a topological phase that is absent in the decoupled limit.
To understand this scenario, we derive low-energy effective Hamiltonians for both the legs.
Note that, for $t_1<t_2$, where the edge modes reside on the leg-a, the effective Hamiltonian of the leg-a is expected to exhibit a nontrivial winding number, while that of the leg-b should remain topologically trivial. 
Conversely, for $t_1>t_2$, where the edge modes are localized on the leg-b, the effective Hamiltonian of the leg-b should display a nontrivial winding, whereas the effective Hamiltonian of the leg-a should not.
In order to confirm this, we derive effective Hamiltonian for both the legs in the limit of weak inter-leg coupling $t_p$ in the following.
\begin{figure}[t]
    \centering \includegraphics[width=1\linewidth]{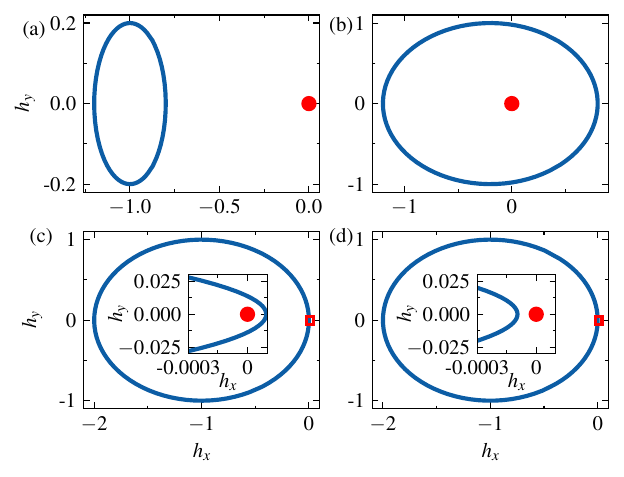}
    \caption{Trajectories of the vector $\vec{h}(k)$ in the $h_x$--$h_y$ plane as $k$ spans the Brillouin zone. Left and right panels correspond to the $t_1>t_2$ and $t_1<t_2$ cases, respectively. Panels (a) and (b) correspond to the effective Hamiltonian of the leg-a obtained by projecting out the leg-b, while panels (c) and (d) correspond to the effective Hamiltonian of the leg-b obtained by projecting out the leg-a. The parameters are $t_p = 0.01$ with (a,c) $t_1=1.0,~ t_2=0.2,~ t=1.0$ and (b,d) $t_1=0.2,~ t_2=1.0,~ t=1.0$. The red rectangular region is magnified in the inset to highlight the winding more clearly.}
    \label{fig:winding}
\end{figure}
To this end, we start from the corresponding momentum space Hamiltonian
\begin{equation}
\begin{pmatrix}
H_{\text{SSH}} & V \\
V & H_{\text{TB}}
\end{pmatrix}
\begin{pmatrix}
\psi_{\text{SSH}} \\
\psi_{\text{TB}}
\end{pmatrix}
= E^{(n)}
\begin{pmatrix}
\psi_{\text{SSH}} \\
\psi_{\text{TB}}
\end{pmatrix},
\end{equation}
where $E^{(n)}$ is the energy of the $n^{th}$ band, $H_{\text{SSH}}=\begin{pmatrix}
0 & \mu_1 \\
\mu_1^* & 0
\end{pmatrix}$, $H_{\text{TB}}=\begin{pmatrix}
0 & \mu_2 \\
\mu_2^* & 0
\end{pmatrix}$, and $V=\begin{pmatrix}
0 & -t_p \\
-t_p & 0
\end{pmatrix}$. The above matrix equation can be written as follows, 
\begin{align}
H_{\text{SSH}}\psi_{\text{SSH}} + V\psi_{\text{TB}} &= E^{(n)}\psi_{\text{SSH}} \\
V\psi_{\text{SSH}} + H_{\text{TB}}\psi_{\text{TB}} &= E^{(n)}\psi_{\text{TB}}
\end{align}
Combining these two equations, we can write,
\begin{equation}
H_{\text{TB}}\psi_{\text{TB}}
+ V \frac{1}{E^{(n)} - H_{\text{SSH}}} V \psi_{\text{TB}}
= E^{(n)}\psi_{\text{TB}}.
\label{eq:eff_ham}
\end{equation}
In the limit of $t_p \ll t, t_1, t_2$, we can approximate the above equation [Eq.~\eqref{eq:eff_ham}] as, $H_{\text{TB}}\psi_{\text{TB}} \simeq E_{\text{TB}}^{(n)}\psi_{\text{TB}}$, where $E_{\text{TB}}^{(n)}$ is the energy of the $n^{th}$ band of the Hamiltonian $H_{\text{TB}}$. Therefore, the effective Hamiltonian of the leg-b in this weak $t_p$ limit looks like,
\begin{equation}
H_{\text{TB}}^{\mathrm{eff}}
= H_{\text{TB}}
+ V \frac{1}{E_{\text{TB}}^{(n)} - H_{\text{SSH}}} V.
\end{equation}
The matrix format of $H_{\text{TB}}^{\mathrm{eff}}$, ignoring the term proportional to the identity matrix, is given by
\begin{align}
H_{\text{TB}}^{\text{eff}}&=
\begin{pmatrix}
    0 & \frac{t_p^2}{{{E_{\text{TB}}^{(n)}}^2}-\mu_1^*\mu_1}\mu_1^*+\mu_2 \\
\frac{t_p^2}{{{E_{\text{TB}}^{(n)}}^2}-\mu_1^*\mu_1}\mu_1+\mu_2^* & 0
\end{pmatrix} \nonumber\\
&=
\begin{pmatrix}
    0 & h_x-ih_y \\
    h_x+ih_y & 0
\end{pmatrix}
=h_x\sigma^x+h_y\sigma^y,
\end{align}
where, $h_x = -(t+\frac{t_p^2}{{E_{\text{TB}}^{(n)}}^2-\mu_1^*\mu_1}t_1)-(t+\frac{t_p^2}{{E_{\text{TB}}^{(n)}}^2-\mu_1^*\mu_1}t_2)\text{cos}(k)$,  $h_y=(t+\frac{t_p^2}{{E_{\text{TB}}^{(n)}}^2-\mu_1^*\mu_1}t_2)\text{sin}(k)$ and $\sigma$'s are the Pauli matrices.
Following a similar method, the effective Hamiltonian of the leg-a is found out to be
\begin{align}
H_{\text{SSH}}^{\text{eff}}&=
\begin{pmatrix}
    0 & \mu_1+\frac{t_p^2}{{{E_{\text{SSH}}^{(n)}}^2}-\mu_2^*\mu_2}\mu_2^* \\
\mu_1^*+\frac{t_p^2}{{{E_{\text{SSH}}^{(n)}}^2}-\mu_2^*\mu_2}\mu_2 & 0
\end{pmatrix}\nonumber\\
&=
\begin{pmatrix}
    0 & h_x-ih_y \\
    h_x+ih_y & 0
\end{pmatrix}
=h_x\sigma^x+h_y\sigma^y,
\label{eq:eff_ssh}
\end{align}
where, $h_x =-(t_1+\frac{t_p^2}{{{E_{\text{SSH}}^{(n)}}^2}-\mu_2^*\mu_2} t)-(t_2+\frac{t_p^2}{{{E_{\text{SSH}}^{(n)}}^2}-\mu_2^*\mu_2}t)\text{cos}(k) $, $h_y = -(t_2+\frac{t_p^2}{{{E_{\text{SSH}}^{(n)}}^2}-\mu_2^*\mu_2}t)\text{sin}(k)$.
The upper and lower panels of Fig.~\ref{fig:winding} show the trajectories of the vector $\vec{h}=(h_x,h_y)$ in the $h_x$–$h_y$ plane for the effective Hamiltonians of leg-a and leg-b, respectively. 
The left and right panels correspond to the cases $t_1>t_2$ and $t_1<t_2$, respectively.
For the effective Hamiltonian of leg-a, the trajectories for $t_p=0.01$ are shown in Fig.~\ref{fig:winding}(a) and (b), corresponding to the parameter sets $(t_1,t_2)=(1.0,0.2)$ and $(0.2,1.0)$, respectively. 
Similarly, the corresponding trajectories for the effective Hamiltonian of leg-b are presented in Fig.~\ref{fig:winding}(c) and (d).
Note that, for this calculation, we avoid the values where ${{E_{\text{TB}}^{(n)}}^2-\mu_1^*\mu_1}$ and ${{E_{\text{SSH}}^{(n)}}^2-\mu_2^*\mu_2}$ become nearly zero.
The topological nature of the effective models is determined by whether the trajectory of $\vec{h}$ encircles the origin, $(h_x,h_y)=(0,0)$, marked by the red circle in Fig.~\ref{fig:winding}. 
For leg-a, shown in Fig.~\ref{fig:winding}(a) and (b), the trajectory winds around the origin only when $t_1<t_2$ [Fig.~\ref{fig:winding}(b)], indicating a nontrivial winding number, consistent with the presence of topological edge states in the effective description of leg-a for this parameter regime. 
In contrast, no winding is observed for $t_1>t_2$ [Fig.~\ref{fig:winding}(a)].
The opposite behaviour is observed for leg-b. 
As shown in Fig.~\ref{fig:winding}(c) and (d), the trajectory exhibits a nontrivial winding when $t_1>t_2$, whereas it does not encircle the origin for $t_1<t_2$. 
Thus, the effective Hamiltonian of leg-b becomes topological in the regime where the effective Hamiltonian of leg-a is topologically trivial.
These results are in complete agreement with the edge-state structure of the full ladder model. 
In particular, they demonstrate that for $t_1>t_2$, where the isolated SSH leg is itself topologically trivial, the inter-leg coupling induces an effective topological phase. 
This confirms the origin of the nontrivial topology observed in the full system and explains the emergence of zero-energy edge states in the leg-b despite the absence of topology in the decoupled limit.
Additionally, this also explains that the topology exists in the full system irrespective of the dimerization pattern, which is consistent with our numerical results.
\begin{figure}[t]
\begin{center}
\includegraphics[width=1\columnwidth]{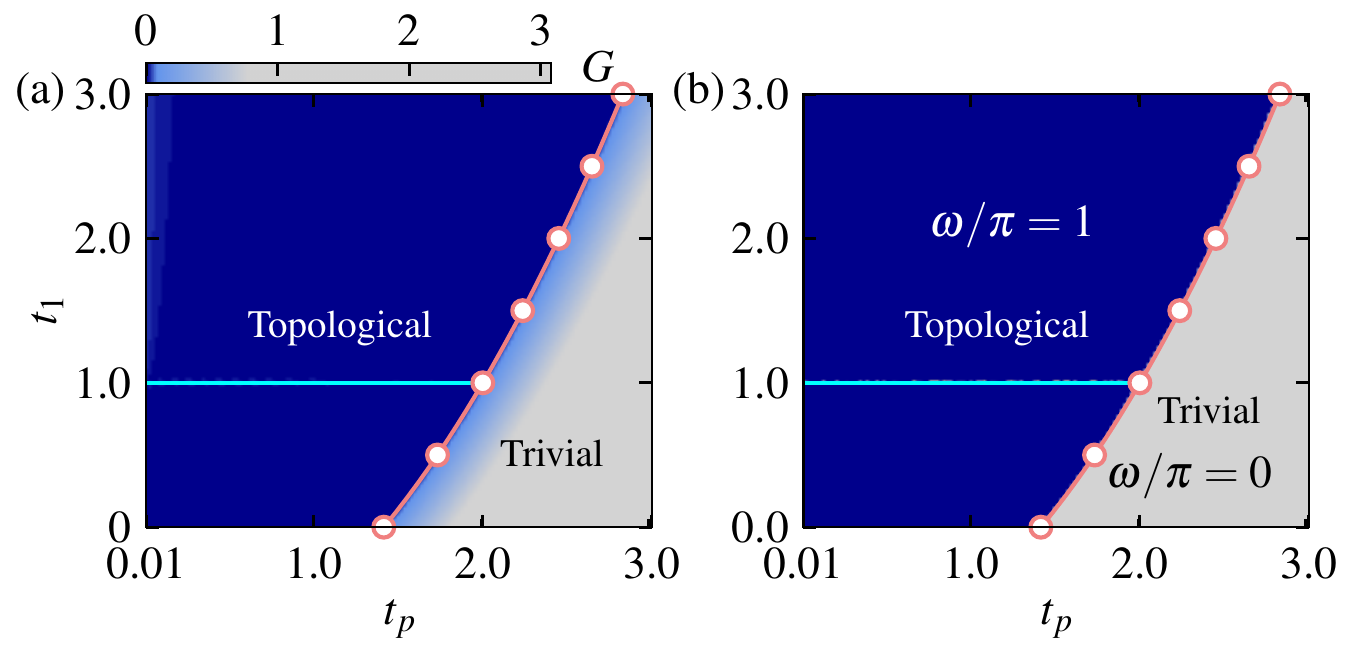}
\end{center}
\caption{The phase diagram in the $t_p-t_1$ plane is shown for $t_2=1.0$. In panel (a), the color bar represents the single-particle excitation gap ($G$) at half-filling, computed for a system of size $L=480$ under OBC. In panel (b), the blue regions correspond to a winding number $\omega/\pi=1$ and gray regions to $\omega/\pi=0$ obtained under PBC. The horizontal lines at $t_1=t_2=1.0$ indicate the gapless line.}
\label{fig:phase_diagram_varying_t1}
\end{figure}
\subsubsection{Phase diagram}
Based on the analysis of the appearance of zero-energy edge modes under OBC and the quantized topological invariant $\omega$ under PBC, we now demonstrate the topological phase transitions in the form of phase diagrams. For this purpose, we characterize the topological properties of the system as a function of $t_p$ for different values of dimerization strength, that is  by fixing $t_2=1.0$ and varying $t_1$.
To this end, we compute the energy gap ($G$), defined as the energy difference between the two eigenstates at the middle of the bulk gap (i.e., at $\rho=1/2$), and is given by $G=E_{L}-E_{L-1}$ under OBC.
The plot of $G$ (color coded) as a function of $t_p$ and $t_1$ while fixing $t_2=1.0$ is shown in  Fig.~\ref{fig:phase_diagram_varying_t1}(a). The blue regions, where $G$ vanishes, indicate the presence of degenerate zero-energy modes at $\rho=1/2$, whereas in the gray region, $G$ acquires finite values due to the absence of any degenerate edge states. 
For a fixed $t_1$, $G$ remains zero up to a critical value of $t_p$, beyond which it becomes finite, suggesting a topological phase transition. 
The point at which this topological phase transition occurs agrees well with the analytical boundary at $t_p=\sqrt{2t(t_1+t_2)}$ (indicated by the red line with circles).
Note that at $t_1=t_2$, the entire bulk is gapless up to $t_p\leq2$ indicated by the horizontal solid line in Fig.~\ref{fig:phase_diagram_varying_t1}(a).
In Fig.~\ref{fig:phase_diagram_varying_t1}(b) we plot the bulk topological invariant $\omega/\pi$ (the winding number) defined in Eq.~\eqref{eq:berry_phase} for the same parameters as in Fig.~\ref{fig:phase_diagram_varying_t1}(a). The values of $\omega/\pi=1~(0)$ confirms the topological (trivial) regions. This clearly complements the topological to trivial phase transition as a function of $t_p$. 
Note that in the gapless region at $t_1=t_2=1.0$ and up to $t_p\leq2$, the invariant is ill-defined due to the absence of a bulk energy gap and hence the value is not fixed in these parameter regimes.

The above analysis demonstrates that a well-defined topological phase emerges in the system due to $t_p$, regardless of the dimerization pattern in leg-a, i.e., whether $t_1<t_2$ or $t_1>t_2$. 
Moreover, an increase in the value of $t_p$ favours a topological to trivial phase transition for any fixed dimerization strength.
In the following, we will quantify the above topological phase transition occurring at $\rho=1/2$ through a quantity that serves as a dynamic probe for characterizing the above topological phase transitions, and also highlight the phenomenon of reversal of Thouless charge pumping.
\begin{figure}[t]
\begin{center}
\includegraphics[width=1.\columnwidth]{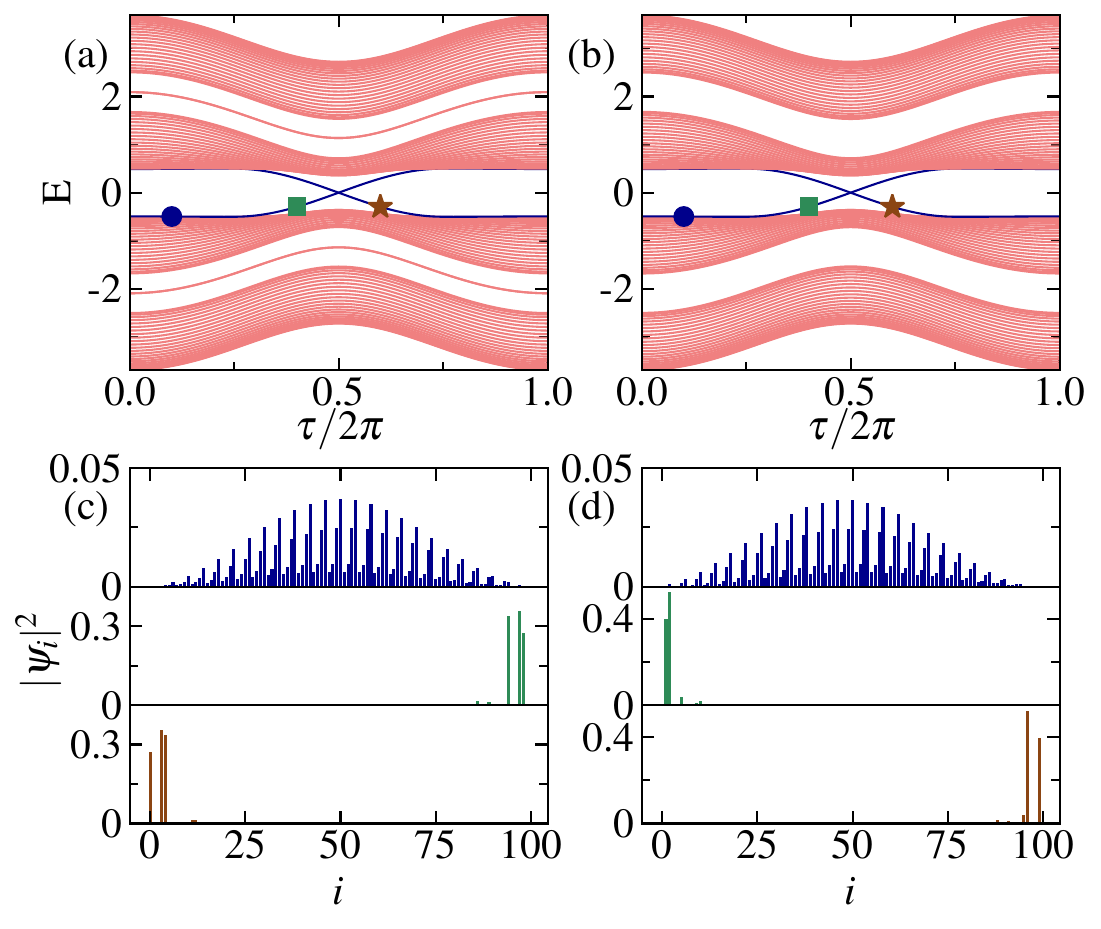}
\end{center}
\caption{Instantaneous energy spectrum of the Hamiltonian $\hat{H}_p$ as a function of the pumping parameter $\tau$, computed for a system of $L = 50$ rungs under OBC. Panel (a) corresponds to the parameter set $t_1 = 0.2$, $t_2 = 1.0$, while panel (b) shows results for $t_1 = 1.0$, $t_2 = 0.2$. In both cases, the pumping parameters are fixed at $t_{p_0} = 1.55$, $\delta_0 = 0.5$, and $\Delta_0 = 0.5$. The highlighted markers in (a) and (b) blue circle, green square, and brown star, indicate specific states whose probability amplitudes are shown in panels (c) and (d), respectively. In panel (c), the top, middle, and bottom subplots show the probability density $|\psi_i|^2$ plotted against the site index for the states marked in (a). Analogously, panel (d) presents the probability density corresponding to the marked states in (b).}
\label{fig:edge_state_rice_mele}
\end{figure}
\subsubsection{Reversal of Thouless charge pumping}
Thouless charge pumping (TCP)~\cite{Thouless1983} is a versatile platform to characterize the topological phase in the system  through periodic modulation of system parameters. This involves slow and cyclic variation of at least two parameters in the system resulting in  a quantized particle transport even in the absence of an external bias.
One of the parameters drives the system through a topological phase transition, while an additional staggered potential term breaks the inversion symmetry and maintains a finite bulk gap throughout the cycle.
This is crucial for ensuring adiabaticity, which is a key requirement for quantized charge transport. 
During the adiabatic cycle, the system traverses from a topological phase to a trivial phase and back to the initial topological phase, enclosing the topological phase transition point. 
This process results in the transport of a quantized number of particles and, the number of particles transported is solely determined by the topological invariant, i.e., the Chern number of the system. 
TCP has been widely investigated in various theoretical and experimental studies in recent years as a key tool for understanding topological phase transitions~\cite{Lohse2016,Takahashi2016pumping,monika_review,Hayward2018,rice_mele,monika_review,Asboth2016_rm,bound_pump1,bound_pump3,juliafare_quantum,Kuno2017,bertok_pump,mondal_phonon,Hayward2018,spin_pumping,pumping_quasicrystals,Taddia2017,pumping_1d,hubbarad_thouless_pump,qubit_pumping,padhan_ladder,seba_nphy_pumming,seba_nature_pumping,pumping_esslinger,pump_reversal_esslinger}.
The Rice-Mele model provides a well-established framework for describing charge pumping protocols. 
To investigate the topological phase transition occurring as a function of $t_p$ at $\rho=1/2$, we propose a pumping protocol that is governed by the Hamiltonian
\begin{align}
    \hat{H}_p &= -\bigg(\sum_{i\in\text{odd}}t_1\hat{a}_i^\dagger\hat{a}_{i+1}+ \sum_{i\in\text{even}}t_2\hat{a}_i^\dagger\hat{a}_{i+1} +t\sum_i\hat{b}_i^\dagger \hat{b}_{i+1} \nonumber \\
    &\;+(t_{p_0}+\delta(\tau))\sum_i\hat{a}_i^\dagger \hat{b}_i\bigg)+\text{H.c.}\nonumber \\
    &\;+ \sum_i \Delta(\tau)((-1)^{i+1} \hat{a}_i^\dagger\hat{a}_i + (-1)^{i}\hat{b}_i^\dagger\hat{b}_i).
    \label{eq:pumping_ham}
\end{align}
Here, $\tau$ is the pumping parameter and $\Delta$ ensures that the bulk gap remains open throughout the cycle.
\begin{figure}[t]
\begin{center}
\includegraphics[width=1.\columnwidth]{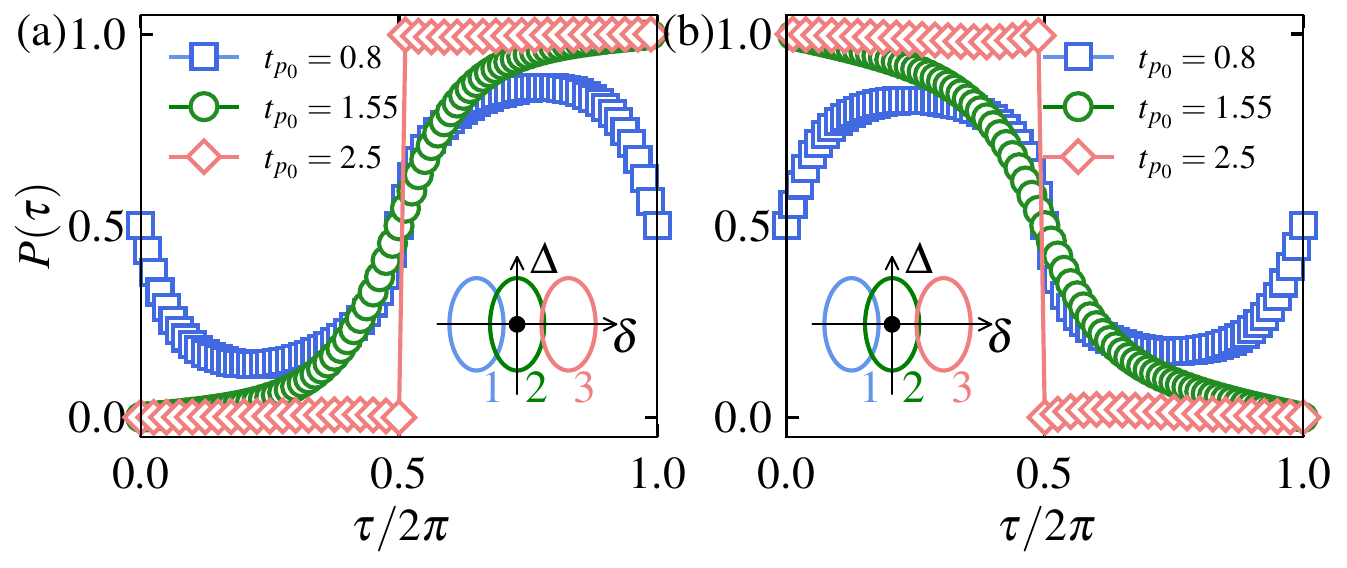}
\end{center}
\caption{Evolution of polarization  $P(\tau)$ as a function of $\tau$ for (a) $t_1=0.2,t_2=1.0$ and (b) $t_1=1.0,t_2=0.2$. The plot shows results for three pumping cycles with different values of $t_{p_0}$ ($=0.8,1.55,2.5$), as indicated in the inset. In all cases, $\delta_0=0.5 ~\text{and}~ \Delta_0=0.5$. The polarization changes are represented by blue squares (cycle 1), green circles (cycle 2), and red diamonds (cycle 3).}
\label{fig:pola}
\end{figure}
($t_{p_0},0$) is the origin of the pumping cycle in the $\delta-\Delta$ plane. 
$\delta(\tau)=\delta_0 \text{cos}(\tau)$ and $\Delta(\tau)=\Delta_0\text{sin}(\tau)$ periodically modulate the rung hopping strength ($t_p$) and the staggered potential term, respectively. 

To investigate the charge pumping mechanism, we present the instantaneous energy spectrum of the Hamiltonian defined in Eq.~\eqref{eq:pumping_ham} for a system of size $L = 100$ rungs under OBC. Fig.~\ref{fig:edge_state_rice_mele}(a) and (b) show the spectra for $t_1 < t_2$ and $t_1 > t_2$, respectively. 
The cyclic variation of the pumping parameters is represented by cycle-2 (highlighted by a green circle in the insets of Fig.~\ref{fig:pola}(a) and (b)), which encloses the topological phase transition point.
In both cases, at $\rho=1/2$, two in-gap modes cross during the cycle, one ascending from the lower to the upper band and the other descending in the opposite direction after a certain value of the pumping parameter. 
To demonstrate their edge nature and the evolution during the pump, we plot the probability density of these modes just before and after the half-cycle (i.e. $\tau/2\pi = 0.5$), which correspond to the green square and brown star marked points in Figs.~\ref{fig:edge_state_rice_mele}(a) and (b), respectively. 
The probability amplitudes are shown in the middle and lower panels of Figs.~\ref{fig:edge_state_rice_mele}(c) and (d), respectively.
These plots confirm that these modes exchange their position from one edge to another edge across the cycle, realizing topological pumping from one edge site to another edge site of the chain. 
For $t_1 < t_2$, the edge mode initially localized at the right end of the chain moves to the left after $\tau/2\pi = 0.5$, and vice versa for $t_1 > t_2$. 
This clearly illustrates the reversal of the topological charge pumping direction, governed by the relative dimerization pattern of $t_1$ and $t_2$.
Since the cycle starts from a trivial phase, initially all states near half-filling have finite probability in the bulk, which is shown in the upper panels of Fig.~\ref{fig:edge_state_rice_mele}(c) and (d) corresponding to the blue circle in Fig.~\ref{fig:edge_state_rice_mele}(a) and (b), respectively.

\begin{figure}[t]
\begin{center}
\includegraphics[width=1.\columnwidth]{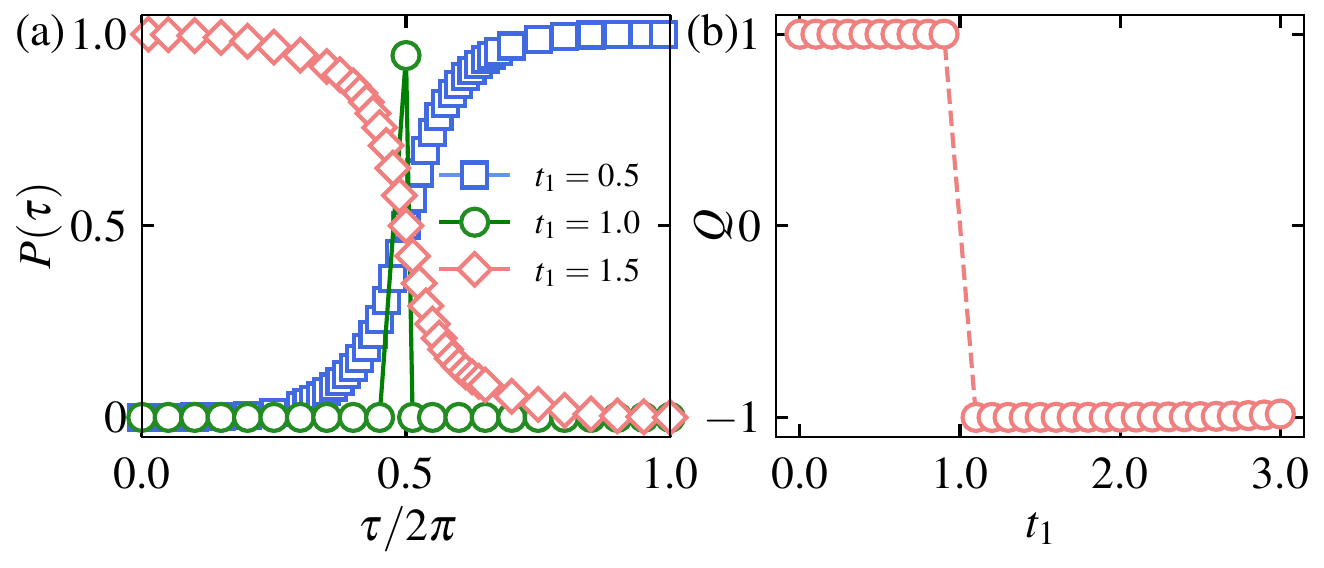}
\end{center}
\caption{(a)The evolution of polarization ($P(\tau)$) as a function of $\tau$ at $t_{p_0}=2$ for three different values of $t_1$ i.e., $0.5$ (blue squares), $1.0$ (green circles), and $1.5$ (red diamonds), keeping $t_2=1$ (Refer Phase diagram shown in Fig.~\ref{fig:phase_diagram_varying_t1}). The pumping parameters are set to be $t_{p_0}=2.0, \delta_0=1.0~~ \text{and},~\Delta_0=0.6$ (b) The amount of charge pumped ($Q$) by varying $t_1$ with fixed value of $t_{p_0}=2.0$.}
\label{fig:rev_pump}
\end{figure}

To complement the reversal of the charge pumping phenomenon, we compute the electronic polarization $P$ using the occupied Bloch bands of the Rice-Mele Hamiltonian under PBC as 
\begin{equation}
P = -\frac{1}{2\pi}\mathfrak{I}\ \ln \prod_{j=0}^{N-1} \text{det}(\langle u_{m,k_j} |u_{n,k_{j+1}} \rangle)~(\text{mod}~Qa),
\label{eq:polarization}
\end{equation}
where $|u_{m,k_j}\rangle$ denotes the occupied Bloch bands of the Hamiltonian 
\begin{equation}
    H_k = \begin{bmatrix}
    \Delta(\tau) & \mu_1 & 0 & -t_p-\delta(\tau)\\
    \mu_1^* & -\Delta(\tau) & -t_p-\delta(\tau) & 0\\
    0 & -t_p-\delta(\tau) & \Delta(\tau) & \mu_2\\
    -t_p-\delta(\tau) & 0 & \mu_2^* & -\Delta(\tau)\\ 
    \end{bmatrix}.
    \label{eq:rice_mele_band_ham}
\end{equation}
Here, $Q$ denotes the particle charge and $a$ is the lattice constant (set to unity). 
We calculate the value of $P$ at different values of $\tau$, which we call $P(\tau)$ and the amount of charge pumped in a cycle is given by
\begin{equation}
 Q = \int_0^{2\pi} d\tau ~\partial_\tau P(\tau).
\label{eq:total_charge}
\end{equation}
It is evident from Fig.~\ref{fig:pola}(a) that only the cycle 2 results in a smooth variation of $P(\tau)$ from $0$ to $1$, corresponding to a quantized charge transfer of $Q = +1$. 
For the cycle-1, no pumping occurs, while the cycle-3 shows discontinuity in $P(\tau)$, indicating a breakdown of transport.
We further explore the scenario by switching the dimerization to $t_1 = 1.0$, $t_2 = 0.2$, while keeping other parameters fixed. 
In this case, $P(\tau)$ evolves from $1$ to $0$ for cycle-2, implying a reversed pumping direction with $Q = -1$ [see Fig.~\ref{fig:pola}(b)]. 
This confirms that the direction of charge transport depends on the dimerization pattern, and altering the dimerization pattern reverses the direction of pumping.
To further illustrate the reversal of charge pumping, we fix the rung hopping to $t_p = 2$ in the phase diagram [Fig.~\ref{fig:phase_diagram_varying_t1}], and examine the evolution of polarization $P(\tau)$ for three different values of $t_1$: 0.5, 1.0, and 1.5, while keeping $t_2 = 1$ fixed [Fig.~\ref{fig:rev_pump}(a)]. 
For $t_1 = 0.5$ ($t_1 < t_2$), the polarization smoothly changes from $0$ to $1$ over the cycle, indicating a pumped charge of $Q = +1$. 
At $t_1 = 1.0$, the pumping path intersects the gap-closing point, leading to a breakdown of adiabatic transport and resulting in no net pumping. 
In contrast, for $t_1 = 1.5$ ($t_1 > t_2$), the polarization evolves from $1$ to $0$, corresponding to a pumped charge of $Q = -1$, demonstrating a reversal in the direction of charge transport compared to the $t_1 < t_2$ case.
To quantify this reversal more generally, we compute the total pumped charge $Q$ as a function of $t_1$, keeping all other parameters fixed as in Fig.~\ref{fig:rev_pump}(a). 
The resulting behavior, shown in Fig.~\ref{fig:rev_pump}(b), clearly shows that the amount of charge pumped changes from $1$ to $-1$ as soon as one crosses the $t_1=t_2$ point. 
This behavior captures the transition from positive to negative charge pumping, confirming the reversal of the charge pumping.
From the above analysis, it is clear that we find the phenomenon of reversal of Thouless charge pumping by periodic modulation of the rung hopping strength.
This phenomenon of reversal of TCP also indicates that the phases for $t_1<t_2$ and $t_1>t_2$ are topologically distinct.

Note that such topological phases do not exist for the hardcore bosonic ladder, and it is solely due to fermionic statistics.

\begin{figure}[t]
\begin{center}
\includegraphics[width=0.85\columnwidth]{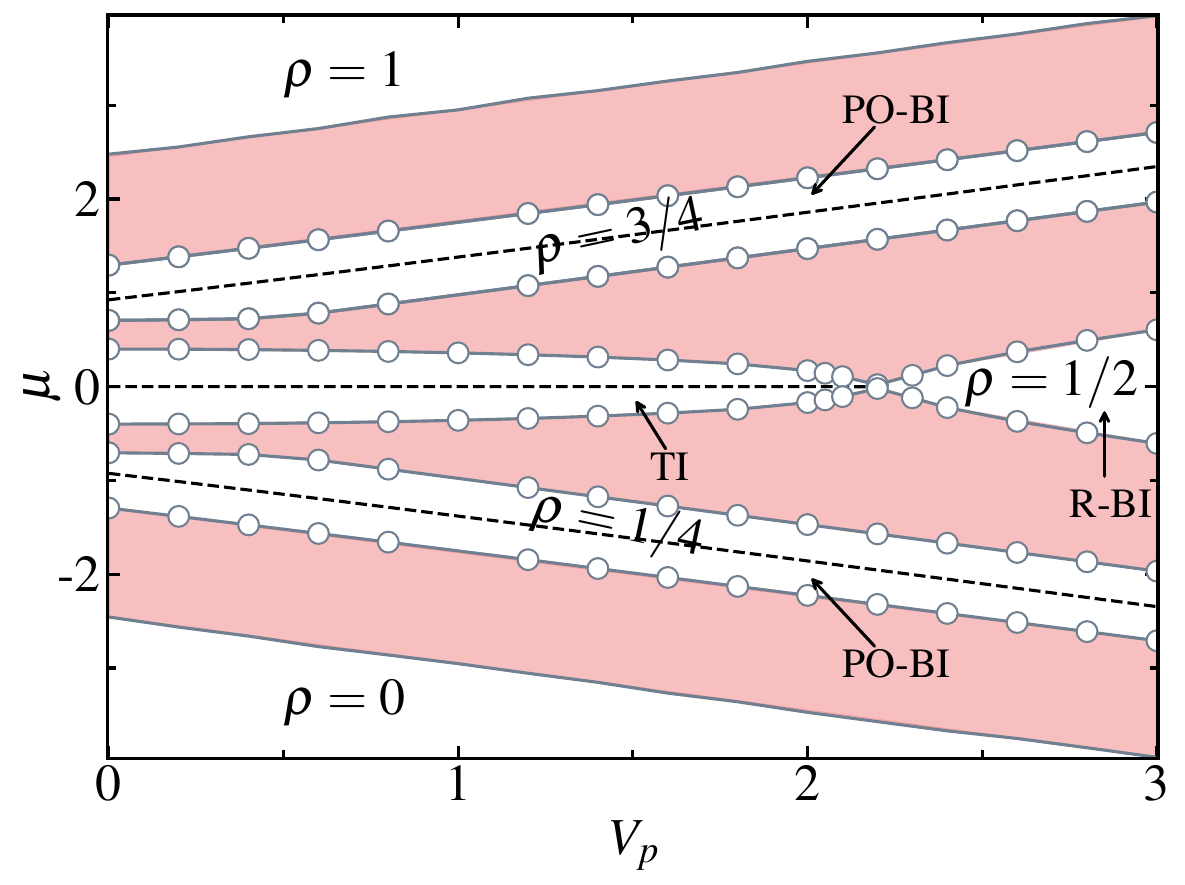}
\end{center}
\caption{The phase diagram in the $V_p-\mu$ plane keeping $t_1=0.2,~t_2=1.0,~t_p=0.8$ fixed. Dashed lines represent the edge states. All the phase boundaries are extrapolated to the thermodynamic limit by taking $L=40,~60,~80,$ and $100$ rungs.}
\label{fig:gap_varying_vp}
\end{figure}

\subsection{Interacting case}\label{many_body}
% \begin{align}
% \hat{H}_M = \hat{H}+\hat{H}_{V_p}
% \label{eq:model_ham}
% \end{align}
We now investigate how this topological phase arising in the non-interacting limit evolves upon introducing interactions in the system. 
Upon introducing interactions between the fermions, the above model exhibits characteristics of a strongly correlated Fermi system. 
For simplicity, we consider only rung-wise repulsive interactions, and hence the system Hamiltonian is now given by, 
\begin{align}
    \hat{H}_M &= \hat{H}+V_p\sum_i(\hat{n}_{i,a}-\frac{1}{2}) (\hat{n}_{i,b}-\frac{1}{2}),
    \label{eq:many_body_ham}
\end{align}
where $\hat{n}_{i,a}$ ($\hat{n}_{i,b}$) represents the occupation density in the upper (lower) leg of the ladder. 
We first explore the effect of inter-chain (rung) repulsive interactions $V_p$ and then the effect of repulsive interactions along all the bonds of the system.

We investigate the many-body ground state properties of the system using the matrix product state (MPS) version of the density matrix renormalization group (DMRG) method~\cite{Verstraete_rev,schollowck_mps} under open boundary conditions (OBC). 
For specific cases, we complement our analysis with exact diagonalization (ED) using periodic boundary conditions (PBC).
In the DMRG calculations, the system size is taken up to $L=200$ rungs ($=400$ sites), with a maximum bond dimension of $800$ to minimize truncation errors.
A system of size $L=12$ rungs ($24$ sites) is used for ED simulations.
Unless otherwise mentioned, all observables are extrapolated to the thermodynamic limit.
\begin{figure}[t]
\begin{center}
\includegraphics[width=0.85\columnwidth]{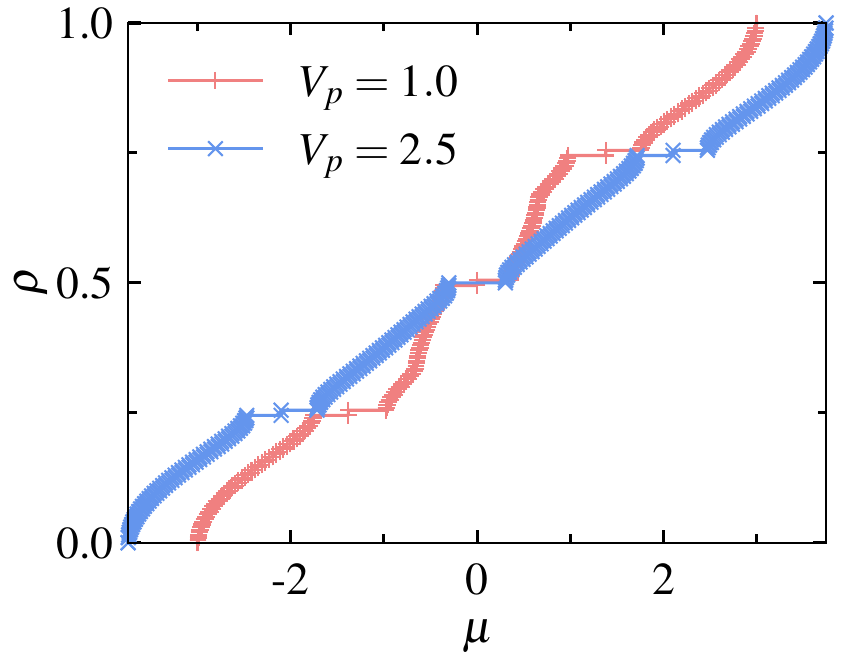}
\end{center}
\caption{The change in particle filling ($\rho$) as a function of change in chemical potential ($\mu$) for different values of $V_p$ for a system of $L=100$ rungs keeping $t_1=0.2,~t_2=1.0,~\text{and}~t_p=0.8$ fixed. The plateaus at different filling ($\rho=1/4,~1/2,~\text{and}~3/4$) indicate the gap in the system and the length of the plateaus give an estimate of the gaps.} \label{fig:rho_mu_gap}
\end{figure}
\subsubsection{Effect of inter-chain interaction}
Before analyzing the evolution of topological properties, we first present the bulk phase diagram in the $V_p-\mu$ plane, with $t_1=0.2,~t_2=1.0$ and $t_p=0.8$, which is depicted in Fig.~\ref{fig:gap_varying_vp}.
The white regions are the gapped phases, and the shaded regions are the gapless phases.
The black dashed lines are the edge states, which are extracted by tracking the vanishing up of the single-particle excitation gap defined as $G=\mu_1^+-\mu_1^-$, where
\begin{equation} \mu^+_1=E_{N+1}-E_N ~~~\text{and}~~~ \mu^-_1=E_N-E_{N-1}. 
\label{eq:gap} \end{equation}
The boundaries of the gapped phases (gray solid lines with circles) are identified from the $\rho$ - $\mu$ plots~\cite{rho_mu_furusaki,rho_mu_tapan,rho_mu_sebastian} for different values of $V_p$. Some exemplary behaviour of $\rho$ as a function of $\mu$ are shown in Fig.~\ref{fig:rho_mu_gap}.

In the $\rho-\mu$ curve, the appearance of plateaus indicates a gap, with the plateau width giving its magnitude.
The lower and the upper phase boundaries shown in Fig.~\ref{fig:gap_varying_vp} are determined from the chemical potential values at the left and right ends of these plateaus in Fig.~\ref{fig:rho_mu_gap}.
All the phase boundaries obtained are extrapolated to the thermodynamic limit.
It is evident from the phase diagram that there are three gapped phases, at $\rho=1/4,~1/2$ and $3/4$ fillings.
The upper (lower) white region is the full (vacuum) state with $\rho=1$ ($\rho=0$).

\begin{figure}[t]
\begin{center}
\includegraphics[width=1.
\columnwidth]{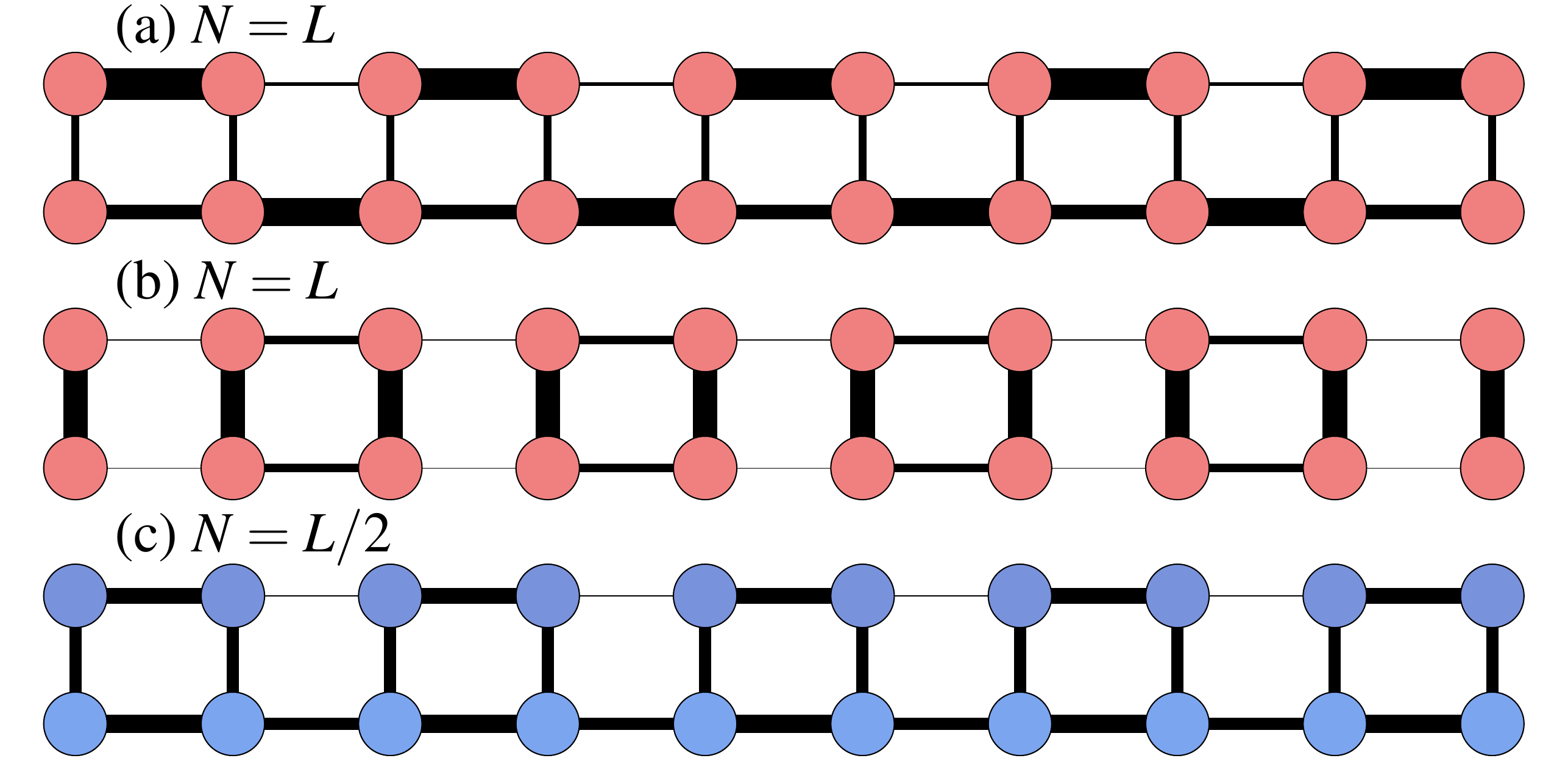}
\end{center}
\caption{The figure shows the magnitude of bond energies of all the bonds defined in Eq.~\eqref{eq:nearest_neighbor_bond_energy} and the on-site particle density for different phases corresponding to Fig. \ref{fig:gap_varying_vp} with a system consisting of $L = 100$ rungs keeping $t_1=0.2,~t_2=1.0,~\text{and}~t_p=0.8$ fixed. In the figures, the thickness of a bond is proportional to the respective magnitude of the bond energies. (a) The magnitude of bond energy and the onsite particle density at $1/2$ filling ($N = L$) for $V_p = 0.2$, (b)  $V_p = 2.8$, and (c) represents for $1/4$ filling ($N=L/2$) at $V_p=1.4$.} \label{fig:many_body_phase_1}
\end{figure}
To gain further insight into the nature of these gapped phases, we analyze the many-body ground state by examining the magnitude of the NN bond energies and onsite particle densities at different fillings $\rho = 1/4,1/2,~\text{and}~3/4$, as shown in the phase diagram in Fig.~\ref{fig:gap_varying_vp}.
To do this, we compute the bond energies on the legs ($B_{i,\alpha}$) and rungs ($B_{i,r}$) using the following expressions:
\begin{equation}
B_{i,\alpha} = \langle\hat{\alpha}_i^\dagger\hat{\alpha}_{i+1} + \text{H.c.}\rangle,\quad
B_{i,r} = \langle \hat{a}_i^\dagger \hat{b}_i + \text{H.c.}\rangle
\label{eq:nearest_neighbor_bond_energy}
\end{equation}
where $\alpha = a, b$ denotes the legs, and $r$ denotes the rungs.
In Fig.~\ref{fig:many_body_phase_1}, the bond thickness represents the magnitude of the bond energy.
The results are obtained for a system of $L=100$ rungs at $V_p = 0.2$, $1.4$, and $2.8$, and the figure focuses only on the central $10$ rungs to highlight the bulk behavior.
At $\rho=1/2$, Fig.~\ref{fig:many_body_phase_1}(a) ($V_p=0.2$) shows strong dimerization along the legs, with alternating bond strengths indicating a bond-ordered (BO) phase. 
In contrast, Fig.~\ref{fig:many_body_phase_1}(b) reveals stronger dimerization along the rungs for $V_p = 2.8$, confirming a rung-dimerized band insulator (R-BI) phase.
The phases at $\rho = 1/4$ and $3/4$ are related by particle-hole symmetry, so we present results only for $\rho = 1/4$ corresponding to $V_p=1.4$.
In this case, a single particle is localized within alternate plaquettes and hop only within them, which we call a plaquette-ordered band insulator (PO-BI) phase.

\begin{figure}[t]
\begin{center}
\includegraphics[width=1.\columnwidth]{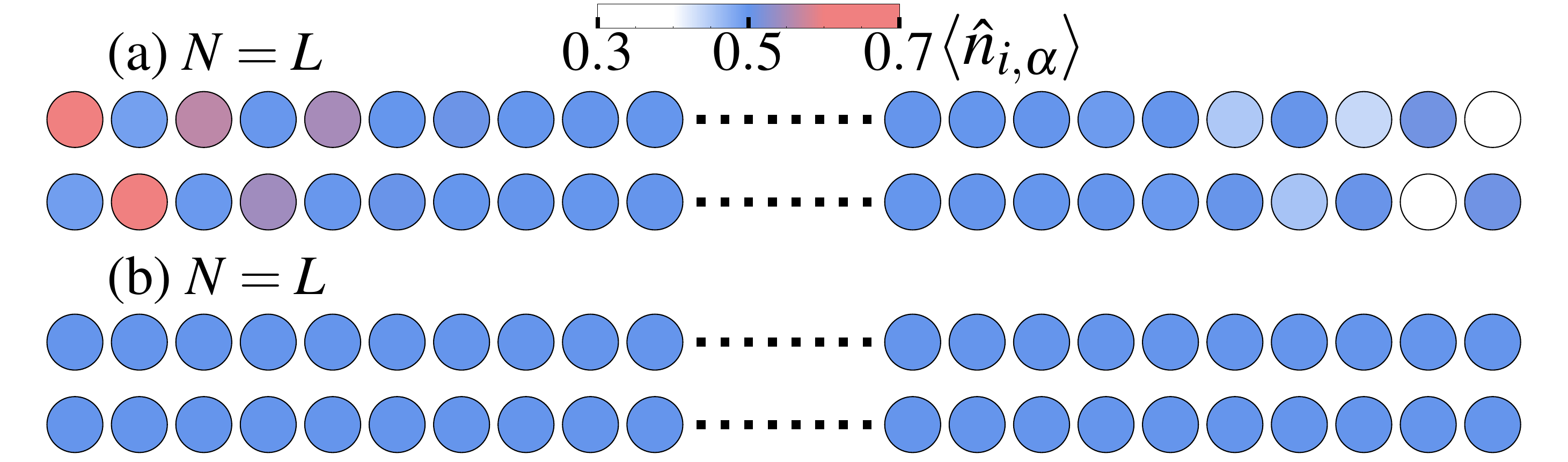}
\end{center}
\caption{On-site particle density for (a) $V_p = 0.2$ and (b) $V_p = 2.8$ keeping $t_1=0.2,~t_2=1.0,~\text{and}~t_p=0.8$ fixed. 
The face color of the circles represents the onsite particle density $\langle \hat{n}_{i,\alpha} \rangle$. 
To highlight the presence of edge states, only a few rungs near the boundary of a ladder with $L = 100$ rungs are displayed.
} \label{fig:many_body_edge_state}
\end{figure}

Building on this understanding of the bulk phases, we now examine how the topological characteristics identified in the non-interacting limit evolve as a function of rung interaction.
Note that the black dashed lines in Fig.~\ref{fig:gap_varying_vp} are the edge states, which also existed in the non-interacting limit (see Fig.~\ref{fig:E_spec}). 
We obtain that these edge states also survive in the presence of interaction and at $\rho=1/2$, with an increase in $V_p$, a gap-closing phase transition occurs to a trivial phase at $V_p\sim2.3$, which we call the interaction-induced topological phase transition.
To show that these states exist at the edges in the presence of finite interaction as well, we plot the on-site particle density of both the legs as a function of the rung index for $V_p=0.2$ in Fig.~\ref{fig:many_body_edge_state}(a). For comparison, we also show the onsite particle density $\langle \hat{n}_{i,\alpha}\rangle$corresponding to the trivial phase, i.e., at $V_p=2.8$, which does not have any edge states as shown in Fig.~\ref{fig:many_body_edge_state}(b).
The clear asymmetry in density observed in both the edge sites for $V_p=0.2$ confirms the edge states.
In contrast, for $V_p=2.8$, the density remains uniform throughout, confirming that there are no edge states in the system.

\begin{figure}[t]
\begin{center}
\includegraphics[width=1\columnwidth]{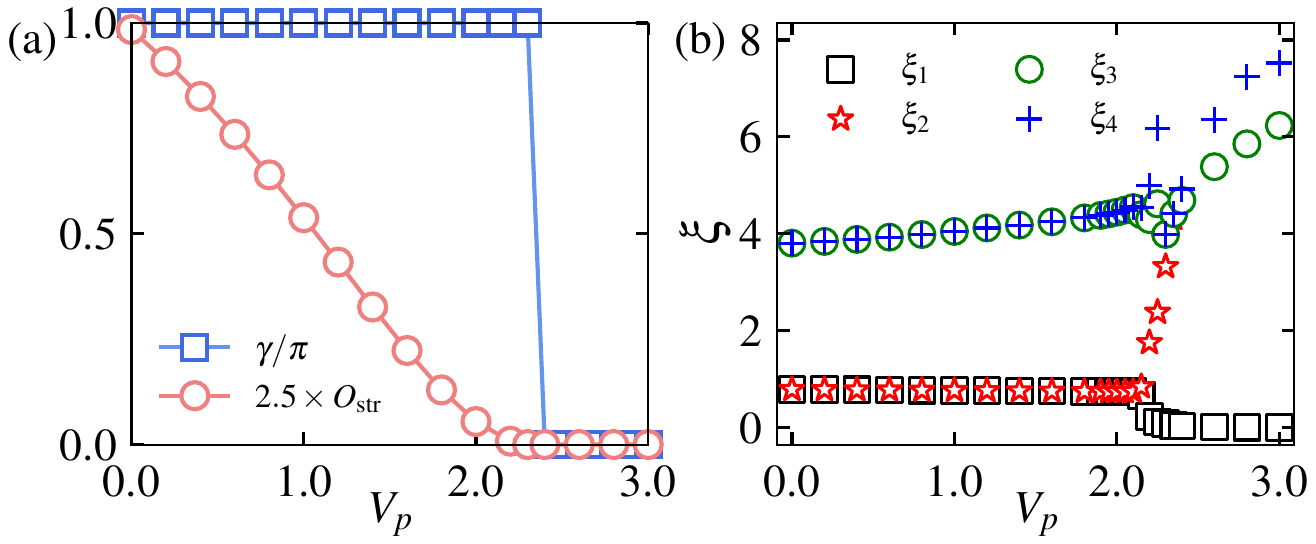}
\end{center}
\caption{(a) shows the winding number $\gamma/\pi$ (squares) calculated using ED for a system of $L=12$ rungs and the string order parameter $O_{\text{str}}$ (circles) computed using DMRG for a system of $L=100$ rungs as a function of $V_p$ with the other parameters set to $t_1=0.2,~t_2=1.0,~\text{and}~t_p=0.8$. (b) shows the lowest four values of the entanglement spectrum ($\xi_1,~\xi_2,~\xi_3,~\xi_4$) as a function of $V_p$, for a system of size $L=200$ rungs.} \label{fig:invariant}
\end{figure}

In order to quantify the topological nature of the system in the presence of interaction, we compute the Berry phase ($\gamma$) for a system of size $L=12$ rungs using ED. The Berry phase $\gamma$, which is regarded as the topological invariant for the interacting system is computed using twisted boundary conditions.
This involves modifying the hopping strength at the boundary as $t \xrightarrow{} te^{i\theta}$, with $\theta$ being the twist angle~\cite{Zak1989,berry_phase_resta,fleishauer_prl}. 
As $\theta$ is varied from $0$ to $2\pi$, the many-body ground state $|\psi(\theta)\rangle$  acquires a geometric phase, known as the Berry phase ($\gamma$)~\cite{berry_phase}, given by
\begin{equation}
\gamma = -\mathfrak{I}\ \ln \prod_{n=0}^{N_\theta-1}\langle \psi_{\theta_n} |\psi_{\theta_{n+1}} \rangle.
\label{eq:berry_phase}
\end{equation}
where $N_\theta$ represents the number of discrete points between $0$ to $2\pi$.
$\gamma$ becomes $0$ in the trivial phase and $\pi$ in the topological phase.
We compute the value of $\gamma/\pi$ known as the winding number, and plot it as a function of $V_p$ in Fig.~\ref{fig:invariant} (a)(blue squares).
The value of $\gamma/\pi$ changes from $1$ to $0$ at $V_p\sim2.3$, confirming the topological phase transition as a function of the interaction $V_p$.

The topological phases are often characterized by non-local string correlations~\cite{den_nijs,Tasaki,Hida,string_ladder}, which can be captured by the string order parameter defined as
\begin{equation}
    O_{\text{str}}(r) = -\langle \hat{Z}_{\text{rung}}(i)e^{i\frac{\pi}{2}\sum_{j=i+1}^{k-1}\hat{Z}_{\text{rung}}(j)}\hat{Z}_{\text{rung}}(k)\rangle,
\label{eq:string_order_parameter}
\end{equation}
where $\hat{Z}_{\text{rung}}(i)=\hat{Z}_a(i)+\hat{Z}_b(i)$ and $\hat{Z}_\alpha(i)=1-2\hat{\alpha}_i^\dagger \hat{\alpha}_i$. 
Here, $i$ denotes the $i^{th}$ rung of the ladder and $r=|i-k|$ represents the distance between the $i^{th}$ and $k^{th}$ rung. 
To avoid edge rungs, we have chosen $i=2$ and $k=L-1$.
We compute $O_{\text{str}}$ for a system of size $L=100$ rungs and plot it along with the winding number in Fig.~\ref{fig:invariant}(a) (red circles).
A comparison between the behavior of $O_{\text{str}}$ and $\gamma/\pi$, reveals that $O_{\text{str}}$ agrees well with the topological phase transition point identified from the winding number.

To further complement this topological phase transition, we compute the entanglement spectrum of the ground state, which serves as a powerful tool for detecting topological order, as it is related to the number of edge excitations in the system~\cite{haldane,kawakami,pollman_entropy,sjia}.
The entanglement spectrum is obtained by logarithmically rescaling the eigenvalues of the reduced density matrix corresponding to a specific bipartition in the system.
For a bipartition at $l^{th}$ bond, the reduced density matrix is defined as $\rho_l = \text{Tr}_{L-l}|\psi\rangle\langle\psi|$, where $|\psi\rangle$ is the ground state wave function.
In the MPS representation used in DMRG, the squared singular values $\Lambda_i^2$ at bond $l$ correspond to the eigenvalues $\lambda_i$ of $\rho_l$. 
The entanglement spectrum is then given by
\begin{equation}
 \xi_i ~=~ - ~\text{ln}(\Lambda_i^2),
\label{eq:entanglement_spectrum}
\end{equation}
We compute the entanglement spectrum at the central bond of a system with $L=200$ rungs and plot the four lowest levels $\xi$ as a function of $V_p$ in Fig.~\ref{fig:invariant}(b). 
A clear two-fold degeneracy appears in the range $0\leq V_p\lesssim 2.3$, which is lifted outside this region.
This degeneracy signals the presence of topological order in the ground state for $0\leq V_p\lesssim 2.3$.

The above analysis demonstrates that the rung interaction ($V_p$) drives a topological phase transition, which in turn enables a Thouless charge pumping mechanism that is modulated by the rung interaction.
Additionally, we also observe that the direction of charge pumping reverses for the opposite dimerization configuration in such an interacting system (not shown).

\begin{figure}[t]
\begin{center}
\includegraphics[width=0.85\columnwidth]{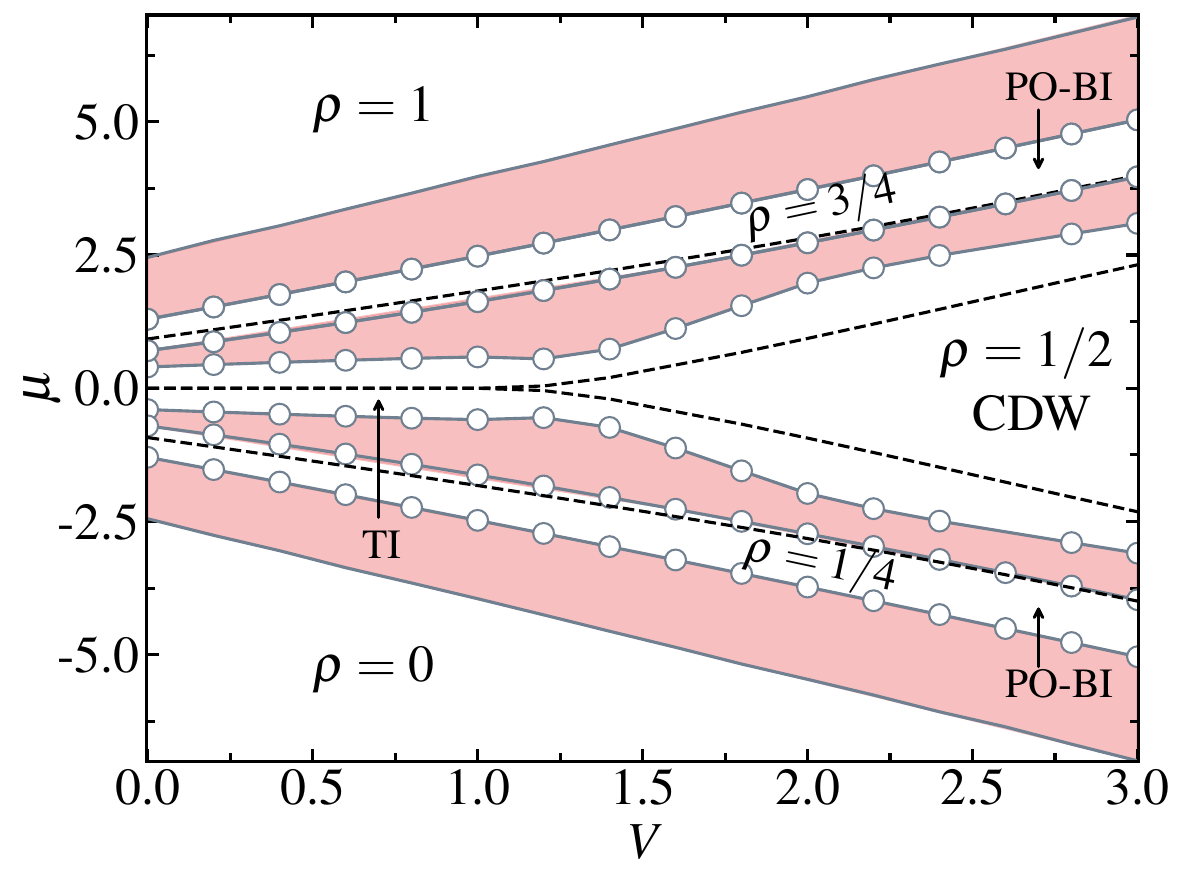}
\end{center}
\caption{The phase diagram in the $V-\mu$ plane keeping $t_1=0.2,~t_2=1.0,~t_p=0.8$ fixed. Dashed lines represent the edge states. All the phase boundaries are extrapolated by taking a maximum system of size $L=100$ rungs.}
\label{fig:gap_varying_uniform_v}
\end{figure}

\subsubsection{Effect of uniform  interaction}\label{uni_int}
For completeness, we also examine the case of uniform NN interactions applied equally along both the legs and the rungs, and analyze their effect on the topological phases present in the non-interacting limit.

\begin{figure}[t]
\begin{center}
\includegraphics[width=1.\columnwidth]{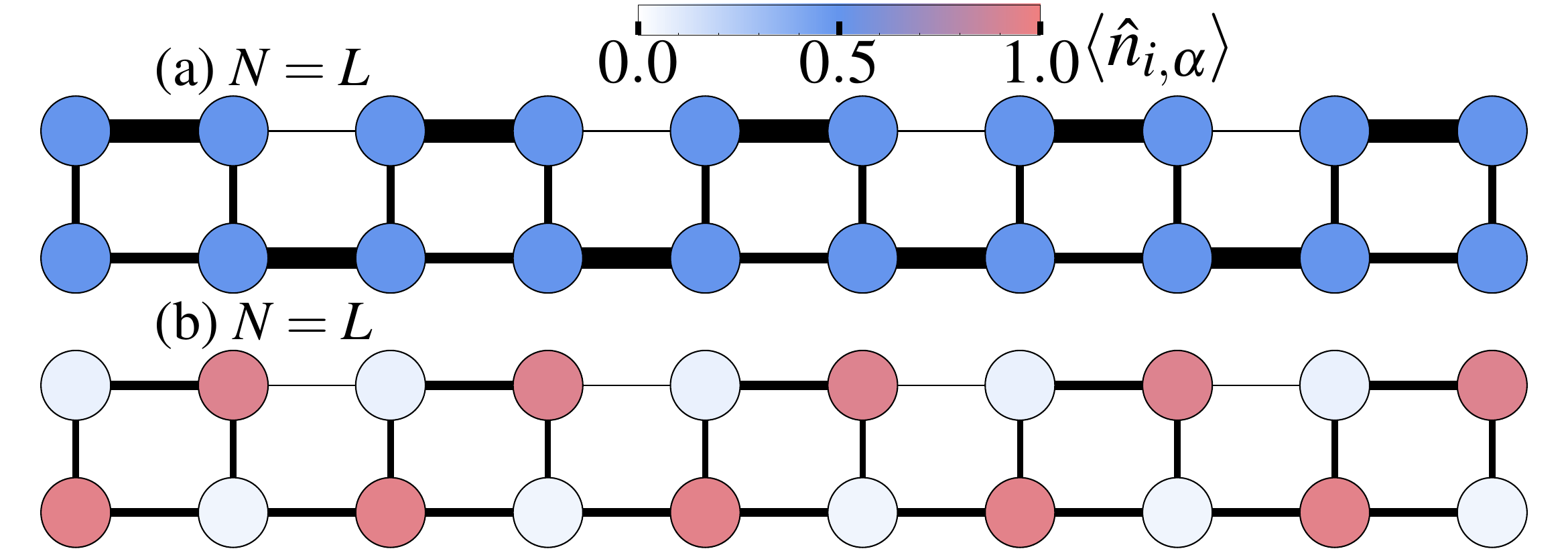}
\end{center}
\caption{The figure shows the magnitude of bond energies of all the bonds defined in Eq.~(\ref{eq:nearest_neighbor_bond_energy}) and the on-site particle density for different phases corresponding to Fig.~\ref{fig:gap_varying_uniform_v} with a system consisting of $L = 100$ rungs keeping $t_1=0.2,~t_2=1.0,~t_p=0.8$ fixed. Only the central $10$ rungs of a system with $L = 100$ rungs are shown to focus on the bulk behavior. In the figures, the thickness of a bond is proportional to the respective magnitude of bond energy, and the face color of the circles represents the values of $\langle \hat{n}_{i,\alpha} \rangle$ for (a) $V = 0.2$, and (b)  $V = 2.8$.} \label{fig:many_body_phases_2}
\end{figure}

The total Hamiltonian in this case is given by
\begin{align}
\hat{H}{_U} = \hat{H} + \hat{H}_{V}
\label{eq:model_ham_uniform_int}
\end{align}
where $\hat{H}$ is the non-interacting part described earlier (Eq.~\eqref{eq:single_particle_ham}), and the interaction term $\hat{H}_{{V}}$ is defined as
\begin{align}
\hat{H}_{V} &= V \sum_i \left(\hat{n}_{i,a} - \frac{1}{2}\right)\left(\hat{n}_{i,b} - \frac{1}{2}\right) \nonumber \\
&\quad + V \sum_{i,\alpha} \left(\hat{n}_{i,\alpha} - \frac{1}{2}\right)\left(\hat{n}_{i+1,\alpha} - \frac{1}{2}\right)
\label{eq:uniform_int}
\end{align}
To study the effect, we begin with a topological phase in the non-interacting limit by setting $t_1 = 0.2$, $t_2 = 1.0$, $t_p = 0.8$, and vary $V$. Figure~\ref{fig:gap_varying_uniform_v} shows the resulting phase diagram in the $V$–$\mu$ plane. 
Three distinct gapped phases emerge at fillings $\rho = 1/4$, $1/2$, and $3/4$.
However, unlike the case with only rung interactions, there is no gap-closing transition at $\rho=1/2$. 
The zero-energy edge modes that exist at $V=0$ (dashed lines) become energetic for $V \gtrsim 1.0$, indicating a transition between two gapped phases.
To characterize the many-body phases at $\rho = 1/2$, we compute the NN bond energies and onsite particle densities as defined in Eq.~\eqref{eq:nearest_neighbor_bond_energy} and depict their magnitude in  Fig.~\ref{fig:many_body_phases_2}(a) and (b) for two representative values of $V=0.2$ and $V=2.8$, respectively.
At $V=0.2$, the prominent oscillation in bond energy with uniform particle densities indicate a BO phase. 
In contrast, for $V=2.8$, the onsite density shows clear oscillations, signaling a charge-density wave (CDW) phase.
Notably, this BO-to-CDW phase transition is not a standard topological phase transition, as it can be described by a local order parameter instead of a global topological invariant.
Thus, we conclude that only rung interactions can drive a true topological phase transition in this system, whereas uniform interactions along both legs and rungs lead to a transition between gapped phases that can be described by local order parameters rather than global topological invariant.

\section{Conclusion}\label{conclusion}
In this study, we have explored the effect of inter-leg coupling in establishing topological phases and phase transitions of spinless fermions on a two-leg ladder. By assuming that one of the legs possessing nearest-neighbour hopping dimerization and the other leg a uniform chain, we have shown that the entire ladder exhibits topological character irrespective of the dimerization pattern in the dimerized leg. 
As a result of this, we obtained two different topological phases of opposite character by changing the dimerization pattern of the leg possessing dimerized hopping. However, stronger inter-leg hopping makes the entire system a trivial insulator.
Additionally, we  showed that the inter-leg interactions can induce a distinct topological phase transition, revealing an interaction-induced topological phase transition in the system. 
We characterized these transitions by multiple diagnostics, including the entanglement spectrum, topological invariants, string order parameters and Thouless charge pumping. 
Our study reveals topological phenomena involving spinless fermions on a quasi-1D geometry, which clearly highlights the intricate interplay between lattice geometry, many-body effects, and particle statistics in stabilizing the topological phase transitions in coupled systems. This also opens up possible avenues to explore the fate of these topological phases in bosonic and spinful fermionic many-body systems, in lattices other than ladder geometries, and in the presence of disorder. 

\section{Acknowledgment}
We thank Nilanjan Roy, Sanchayan Banerjee, and Sitaram Maity for useful discussions. T.M. acknowledges support from Science and Engineering Research Board (SERB), Govt. of India, through project No. MTR/2022/000382 and STR/2022/000023.

\section{Data availability}
The data that support the findings of this article are openly available~\cite{data}.
\bibliography{references.bib}
\end{document}